\pdfoutput=1
\documentclass[12pt,a4paper]{article}

\usepackage{ifthen} \newboolean{pdflatex}
\setboolean{pdflatex}{true} 

\newboolean{articletitles}
\setboolean{articletitles}{true} 

\newboolean{uprightparticles}
\setboolean{uprightparticles}{false}

\def\paperauthors{LHCb collaboration} \def\paperasciititle{Search for lepton-number-violating B- to D(*)+mu-mu- decays} \def\papertitle{\boldmath Search for lepton-number-violating \decay{\Bm}{
\dordstmumu} decays} \def\paperkeywords{{High Energy Physics}, {LHCb}} \def\papercopyright{\the\year\ CERN for the benefit of the LHCb
collaboration} \def\paperlicence{CC BY 4.0 licence}
\def\paperlicenceurl{https://creativecommons.org/licenses/by/4.0/}

\newif\ifEnableSectionTOCLinks
\EnableSectionTOCLinksfalse

\usepackage[top=1in, bottom=1.25in, left=1in, right=1in]{geometry}

\usepackage{microtype}
\usepackage{lineno}  \usepackage{xspace} \usepackage{caption} 

\usepackage{rotating}

\usepackage{graphicx}  \usepackage{color}
\usepackage{colortbl}
\graphicspath{{./figs/}} 

\usepackage{amsmath} \usepackage{amssymb}
\usepackage{amsfonts}
\usepackage{upgreek} 

\newcommand*\patchAmsMathEnvironmentForLineno[1]{\expandafter\let\csname old#1\expandafter\endcsname\csname #1\endcsname
  \expandafter\let\csname oldend#1\expandafter\endcsname\csname
  end#1\endcsname
  \renewenvironment{#1}{\linenomath\csname old#1\endcsname}{\csname oldend#1\endcsname\endlinenomath}}
\newcommand*\patchBothAmsMathEnvironmentsForLineno[1]{\patchAmsMathEnvironmentForLineno{#1}\patchAmsMathEnvironmentForLineno{#1*}}
\AtBeginDocument{\patchBothAmsMathEnvironmentsForLineno{equation}\patchBothAmsMathEnvironmentsForLineno{align}\patchBothAmsMathEnvironmentsForLineno{flalign}\patchBothAmsMathEnvironmentsForLineno{alignat}\patchBothAmsMathEnvironmentsForLineno{gather}\patchBothAmsMathEnvironmentsForLineno{multline}\patchBothAmsMathEnvironmentsForLineno{eqnarray}}

\usepackage[pdftex,
  pdfauthor={\paperauthors},
  pdftitle={\paperasciititle},
pdfkeywords={\paperkeywords}]{hyperref}
\usepackage{hyperxmp}
\hypersetup{
  pdfcopyright={Copyright (C) \papercopyright},
  pdflicenseurl={\paperlicenceurl}
}

\usepackage[colorinlistoftodos,textsize=scriptsize]{todonotes}

\usepackage[bottom,flushmargin,hang,multiple]{footmisc}

\usepackage[all]{hypcap} 

\usepackage{xspace}
\usepackage{upgreek}

\def\lhcb   {\mbox{LHCb}\xspace}

\def\belle  {\mbox{Belle}\xspace}

\def\MagUp {\mbox{\em Mag\kern -0.05em Up}\xspace}

\ifthenelse{\boolean{uprightparticles}}{

 \def\Pmu         {\ensuremath{\upmu}\xspace}
 \def\Pnu         {\ensuremath{\upnu}\xspace}
 
 \def\Ppi         {\ensuremath{\uppi}\xspace}

 \def\Ppsi        {\ensuremath{\uppsi}\xspace}

 \def\PDelta      {\ensuremath{\Delta}\xspace}
 \def\PXi         {\ensuremath{\Xi}\xspace}
 \def\PLambda     {\ensuremath{\Lambda}\xspace}
 \def\PSigma      {\ensuremath{\Sigma}\xspace}
 \def\POmega      {\ensuremath{\Omega}\xspace}
 \def\PUpsilon    {\ensuremath{\Upsilon}\xspace}
 \let\oldPi\Pi
 \def\PPi         {\ensuremath{\oldPi}\xspace}

\def\PB      {\ensuremath{\mathrm{B}}\xspace}
\def\PD      {\ensuremath{\mathrm{D}}\xspace}

\def\PJ      {\ensuremath{\mathrm{J}}\xspace}
 \def\PK      {\ensuremath{\mathrm{K}}\xspace}

\def\Pb      {\ensuremath{\mathrm{b}}\xspace}
 \def\Pc      {\ensuremath{\mathrm{c}}\xspace}

\def\Pp      {\ensuremath{\mathrm{p}}\xspace}
 
\def\Ps      {\ensuremath{\mathrm{s}}\xspace}

\def\thebaroffset{0.0em}
}
{

 \def\Pmu         {\ensuremath{\mu}\xspace}
 \def\Pnu         {\ensuremath{\nu}\xspace}
 
 \def\Ppi         {\ensuremath{\pi}\xspace}

 \def\Ppsi        {\ensuremath{\psi}\xspace}
 
 \mathchardef\PDelta="7101
 \mathchardef\PXi="7104
 \mathchardef\PLambda="7103
 \mathchardef\PSigma="7106
 \mathchardef\POmega="710A
 \mathchardef\PUpsilon="7107
 \mathchardef\PPi="7105
\def\PB      {\ensuremath{B}\xspace}
\def\PD      {\ensuremath{D}\xspace}

\def\PJ      {\ensuremath{J}\xspace}
 \def\PK      {\ensuremath{K}\xspace}

\def\Pb      {\ensuremath{b}\xspace}
 \def\Pc      {\ensuremath{c}\xspace}

\def\Pp      {\ensuremath{p}\xspace}
 
\def\Ps      {\ensuremath{s}\xspace}

\def\thebaroffset{0.18em}
}
\newcommand{\offsetoverline}[2][\thebaroffset]{\kern #1\overline{\kern -#1 #2}}

\makeatletter
\ifcase \@ptsize \relax \newcommand{\miniscule}{\@setfontsize\miniscule{4}{5}}\or \newcommand{\miniscule}{\@setfontsize\miniscule{5}{6}}\or \newcommand{\miniscule}{\@setfontsize\miniscule{5}{6}}\fi
\makeatother

\DeclareRobustCommand{\optbar}[1]{\shortstack{{\miniscule (\rule[.5ex]{1.25em}{.18mm})}
  \\ [-.7ex] $#1$}}

\def\mup        {{\ensuremath{\Pmu^+}}\xspace}
\def\mun        {{\ensuremath{\Pmu^-}}\xspace}

\def\neu        {{\ensuremath{\Pnu}}\xspace}
\def\neub       {{\ensuremath{\overline{\Pnu}}}\xspace}

\def\neum       {{\ensuremath{\neu_\mu}}\xspace}
\def\neumb      {{\ensuremath{\neub_\mu}}\xspace}

\def\squark    {{\ensuremath{\Ps}}\xspace}

\def\cquark    {{\ensuremath{\Pc}}\xspace}

\def\bquark    {{\ensuremath{\Pb}}\xspace}

\def\pion   {{\ensuremath{\Ppi}}\xspace}

\def\pip    {{\ensuremath{\pion^+}}\xspace}
\def\pim    {{\ensuremath{\pion^-}}\xspace}

\def\kaon    {{\ensuremath{\PK}}\xspace}

\def\KorKbar {\kern \thebaroffset\optbar{\kern -\thebaroffset \PK}{}\xspace}

\def\Km      {{\ensuremath{\kaon^-}}\xspace}

\def\D       {{\ensuremath{\PD}}\xspace}

\def\DorDbar {\kern \thebaroffset\optbar{\kern -\thebaroffset \PD}\xspace}
\def\Dz      {{\ensuremath{\D^0}}\xspace}

\def\Dp      {{\ensuremath{\D^+}}\xspace}
\def\Dm      {{\ensuremath{\D^-}}\xspace}

\def\DpDm    {\ensuremath{\Dp {\kern -0.16em \Dm}}\xspace}

\def\Dstarp  {{\ensuremath{\D^{*+}}}\xspace}

\def\B       {{\ensuremath{\PB}}\xspace}

\def\BorBbar {\kern \thebaroffset\optbar{\kern -\thebaroffset \PB}\xspace}

\def\Bd      {{\ensuremath{\B^0}}\xspace}

\def\BdorBdbar {\kern \thebaroffset\optbar{\kern -\thebaroffset \Bd}\xspace}

\def\Bub     {{\ensuremath{\B^-}}\xspace}

\def\Bm      {{\ensuremath{\Bub}}\xspace}

\def\Bs      {{\ensuremath{\B^0_\squark}}\xspace}

\def\BsorBsbar {\kern \thebaroffset\optbar{\kern -\thebaroffset \Bs}\xspace}

\def\jpsi     {{\ensuremath{{\PJ\mskip -3mu/\mskip -2mu\Ppsi}}}\xspace}
\def\psitwos  {{\ensuremath{\Ppsi{(2S)}}}\xspace}

\def\Y#1S{\ensuremath{\PUpsilon{(#1S)}}\xspace}

\def\proton      {{\ensuremath{\Pp}}\xspace}

\def\LorLbar     {\kern \thebaroffset\optbar{\kern -\thebaroffset \PLambda}\xspace}

\def\BF         {{\ensuremath{\mathcal{B}}}\xspace}

\newcommand{\decay}[2]{\mbox{\ensuremath{#1\!\to #2}}\xspace}

\def\to                 {\ensuremath{\rightarrow}\xspace}

\def\order   {{\ensuremath{\mathcal{O}}}\xspace}

\def\eps   {{\ensuremath{\varepsilon}}\xspace}

\def\AT#1     {\ensuremath{A_{\mathrm{T}}^{#1}}\xspace}

\def\C#1      {\ensuremath{\mathcal{C}_{#1}}\xspace}                       \def\Cp#1     {\ensuremath{\mathcal{C}_{#1}^{'}}\xspace}                    \def\Ceff#1   {\ensuremath{\mathcal{C}_{#1}^{\mathrm{(eff)}}}\xspace}        \def\Cpeff#1  {\ensuremath{\mathcal{C}_{#1}^{'\mathrm{(eff)}}}\xspace}       \def\Ope#1    {\ensuremath{\mathcal{O}_{#1}}\xspace}                       \def\Opep#1   {\ensuremath{\mathcal{O}_{#1}^{'}}\xspace}

\newcommand{\nospaceunit}[1]{\ensuremath{\text{#1}}}
\newcommand{\aunit}[1]{\ensuremath{\text{\,#1}}}

\newcommand{\tev}{\aunit{Te\kern -0.1em V}\xspace}
\newcommand{\gev}{\aunit{Ge\kern -0.1em V}\xspace}
\newcommand{\mev}{\aunit{Me\kern -0.1em V}\xspace}
\newcommand{\kev}{\aunit{ke\kern -0.1em V}\xspace}
\newcommand{\ev}{\aunit{e\kern -0.1em V}\xspace}

\newcommand{\mevc}{\ensuremath{\aunit{Me\kern -0.1em V\!/}c}\xspace}
\newcommand{\gevc}{\ensuremath{\aunit{Ge\kern -0.1em V\!/}c}\xspace}
\newcommand{\mevcc}{\ensuremath{\aunit{Me\kern -0.1em V\!/}c^2}\xspace}
\newcommand{\gevcc}{\ensuremath{\aunit{Ge\kern -0.1em V\!/}c^2}\xspace}

\def\mum  {\ensuremath{\,\upmu\nospaceunit{m}}\xspace}

\def\fb   {\ensuremath{\aunit{fb}}\xspace}
\def\invfb   {\ensuremath{\fb^{-1}}\xspace}

\def\order{{\ensuremath{\mathcal{O}}}\xspace}
\newcommand{\chisq}{\ensuremath{\chi^2}\xspace}

\newcommand{\chisqip}{\ensuremath{\chi^2_{\text{IP}}}\xspace}

\def\gsim{{~\raise.15em\hbox{$>$}\kern-.85em
          \lower.35em\hbox{$\sim$}~}\xspace}
\def\lsim{{~\raise.15em\hbox{$<$}\kern-.85em
          \lower.35em\hbox{$\sim$}~}\xspace}

\def\sPlot{\mbox{\em sPlot}\xspace}

\def\pt         {\ensuremath{p_{\mathrm{T}}}\xspace}

\def\ptot       {\ensuremath{p}\xspace}

\def\evtgen     {\mbox{\textsc{EvtGen}}\xspace}

\def\geant      {\mbox{\textsc{Geant4}}\xspace}

\def\photos     {\mbox{\textsc{Photos}}\xspace}

\def\pythia     {\mbox{\textsc{Pythia}}\xspace}

\def\tell1  {TELL1\xspace}
\def\ukl1   {UKL1\xspace}

\newcommand{\lhcborcid}[1]{\href{https://orcid.org/#1}{\hspace*{0.1em}\raisebox{-0.45ex}{\includegraphics[width=1em]{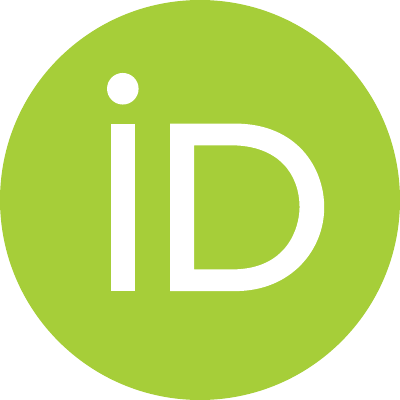}}}}

 \def\db {\ensuremath{0\nu\beta\beta}\xspace}

\def\dordstpipi{\ensuremath{D^{(*)+}\pi^-\pi^-}\xspace}

\def\dordstmumu{\ensuremath{D^{(*)+}\mu^-\mu^-}\xspace}

\def\dordstarp  {{\ensuremath{\D^{(*)+}}}\xspace}

\def\roostat     {\mbox{\textsc{RooStats}}\xspace}
 
\hypersetup{
  colorlinks   = true, urlcolor     = blue, linkcolor    = blue, citecolor    = red   }

\ifEnableSectionTOCLinks
\usepackage[explicit]{titlesec} 

\let\oldcontentsline\contentsline
\renewcommand

\titleformat{\section}{\normalfont\Large\bf}{\hyperlink{tocsection.\thesection}{{\thesection}
\parbox[t]{\dimexpr\textwidth-1pc}{#1}}}{1pc}{}

\titleformat{\subsection}{\normalfont\bf}{\hyperlink{tocsubsection.\thesubsection}{{\thesubsection}
\parbox[t]{\dimexpr\textwidth-1pc}{#1}}}{1pc}{}

\titleformat{name=\section,numberless}[display]{}{}{0pt}{\normalfont\Huge\bfseries
#1}
\fi

\usepackage{cite} \usepackage{mciteplus}
 \usepackage{booktabs} \usepackage[compat=1.1.0]{tikz-feynman}
\usepackage{longtable} 

\begin{document}

\renewcommand{\thefootnote}{\fnsymbol{footnote}}
\setcounter{footnote}{1}

\begin{titlepage}
  \pagenumbering{roman}

\vspace*{-1.5cm}
  \centerline{\large EUROPEAN ORGANIZATION FOR NUCLEAR RESEARCH (CERN)}
  \vspace*{1.5cm}
  \noindent
  \begin{tabular*}{\linewidth}{lc@{\extracolsep{\fill}}r@{\extracolsep{0pt}}}
    \ifthenelse{\boolean{pdflatex}}{\vspace*{-1.5cm}\mbox{\!\!\!\includegraphics[width=.14\textwidth]{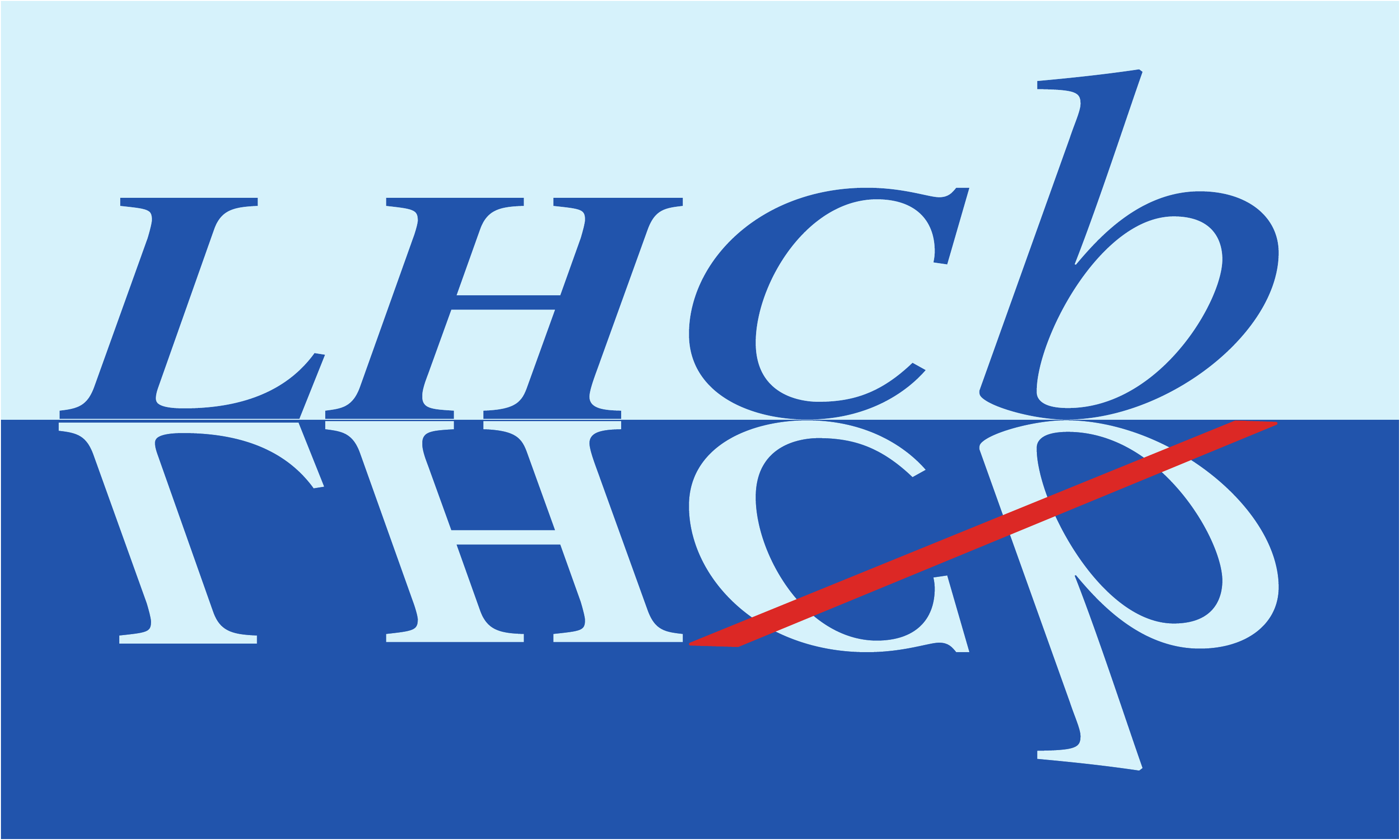}}
    & &}{\vspace*{-1.2cm}\mbox{\!\!\!\includegraphics[width=.12\textwidth]{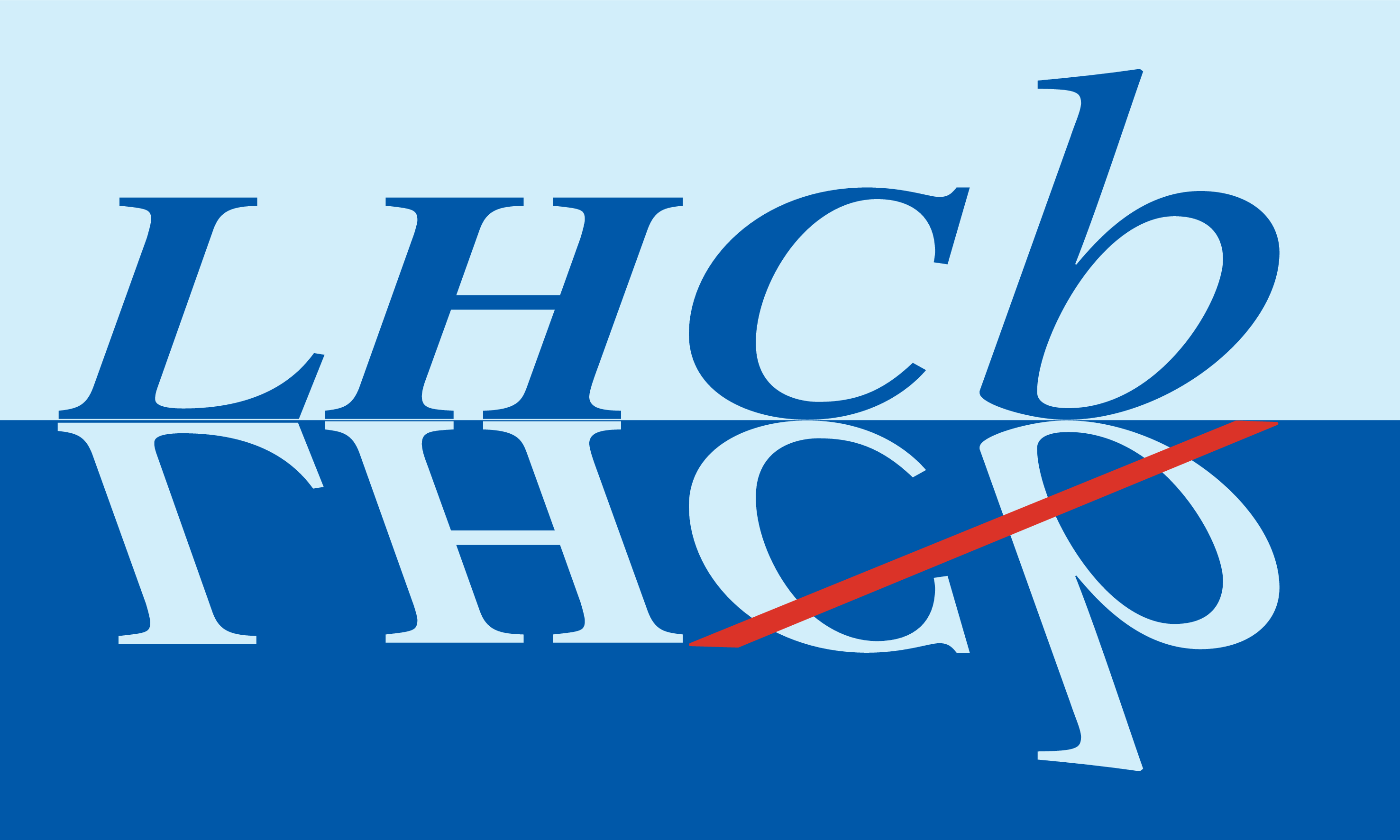}}
    & &}\\
    & & CERN-EP-2025-267 \\  & & LHCb-PAPER-2025-033 \\  & & May 26, 2026 \\ & & \\
\end{tabular*}

  \vspace*{4.0cm}

{\normalfont\bfseries\boldmath\huge
    \begin{center}
\papertitle
    \end{center}
  }

  \vspace*{2.0cm}

\begin{center}
\paperauthors\footnote{Authors are listed at the end of this paper.}
  \end{center}

  \vspace{\fill}

\begin{abstract}
    \noindent
    A search is performed for lepton-number-violating
    \decay{\Bm}{\dordstmumu} decays, using data
    collected by the LHCb experiment in proton-proton collisions at a
    center-of-mass energy of 13\tev, corresponding to an integrated luminosity of 5.4\invfb. No significant signal is observed, and 
    upper limits are set on the branching fractions, \mbox{$\BF (\decay{\Bm}{\Dp\mun\mun}) < 4.6 \times 10^{-8}$} and
	\mbox{$\BF (\decay{\Bm}{\Dstarp\mun\mun}) < 5.9 \times 10^{-8}$}, at the 95\% confidence level.
  \end{abstract}

  \vspace*{2.0cm}

  \begin{center}
    Published in
    Physical Review D 113, 092006 (2026)
  \end{center}

  \vspace{\fill}

  {\footnotesize
\centerline{\copyright~\papercopyright.
  \href{\paperlicenceurl}{\paperlicence}.}}
  \vspace*{2mm}

\end{titlepage}

\newpage
\setcounter{page}{2}
\mbox{~}

\renewcommand{\thefootnote}{\arabic{footnote}}
\setcounter{footnote}{0}

\cleardoublepage

\pagestyle{plain} \setcounter{page}{1}
\pagenumbering{arabic}

\section{Introduction}
\label{sec:Introduction}

\begin{figure}[b]
  \centering
  \includegraphics[width=0.5\textwidth]{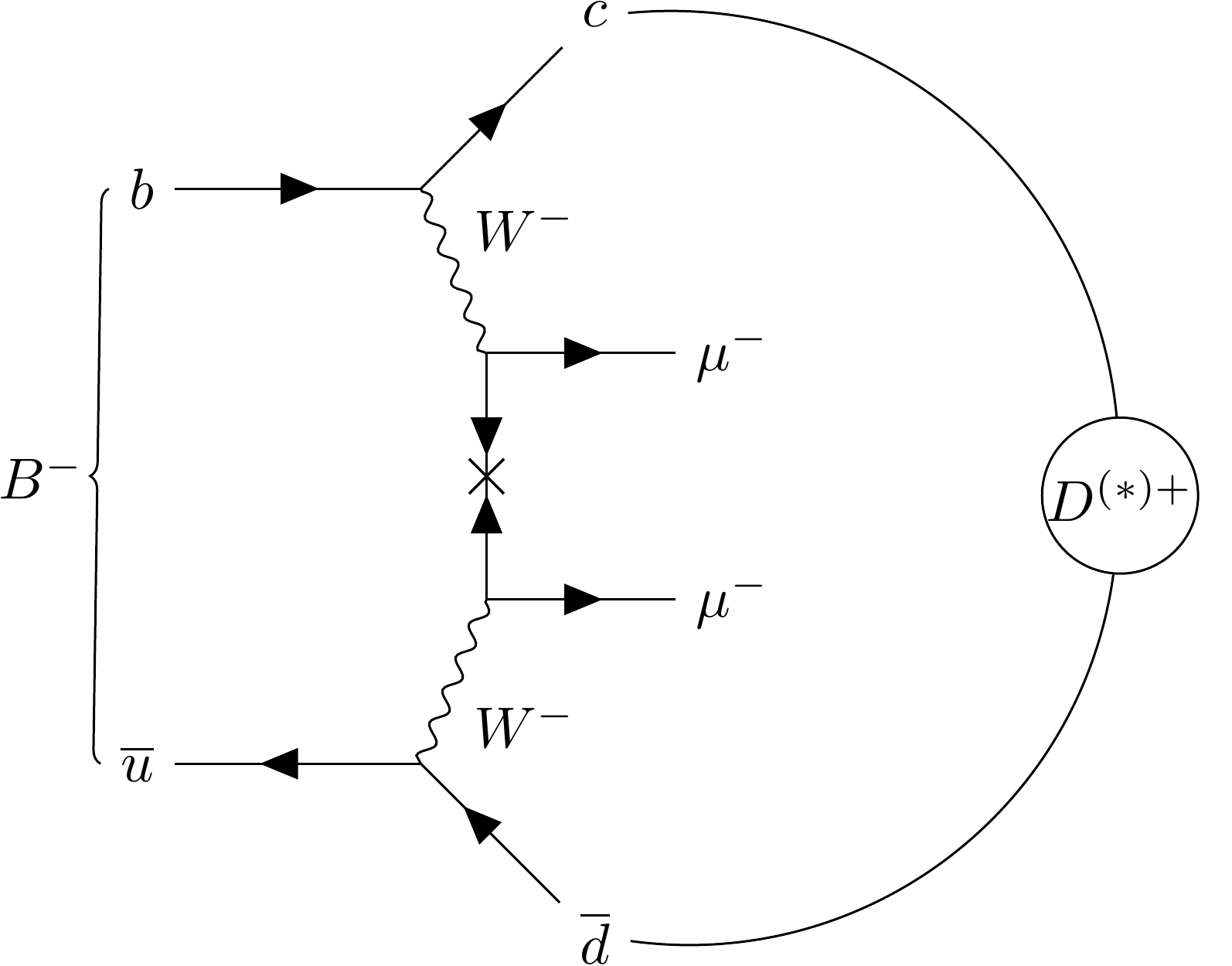}
  \caption{Diagram for the $b\to u$ annihilation process. The ``$\times$"
  indicates the $\nu_\mu-\overline{\nu}_\mu$ exchange.}
  \label{fig:0vbb}
\end{figure}

One of the fundamental questions in particle physics is whether neutrinos are Dirac or Majorana fermions.
Dirac fermions have distinct particle and antiparticle states, whereas Majorana fermions are identical to their antiparticles.
A key way to establish the Majorana nature of neutrinos is through the observation of neutrinoless double beta decay (\db), a second-order weak process in which two neutrons in the same nucleus undergo correlated $\beta^-$ transitions, emitting two electrons but no neutrinos.
In the Standard Model (SM), only the two-neutrino double beta decay process can occur, in which the final state contains two electrons and two antineutrinos.
The most stringent lower limits on \db lifetimes have been set by the LEGEND-200 experiment, $\tau  > 1.9\times 10^{26}$~years for $^{76}$Ge nuclei~\cite{Romo-Luque:2025skd}, and by the EXO experiment, $\tau > 3.5\times 10^{25}$~years for $^{136}$Xe nuclei~\cite{EXO-200:2019rkq}.

While the \db experiments can only probe the couplings of Majorana neutrinos to electrons, the couplings to muons can be investigated using analogous
$s$-, $c$-, or $b$-hadron decays. The LHCb and Belle experiments have performed searches with \Bm mesons~\cite{BELLE:2011bej,LHCb-PAPER-2011-038,LHCb-PAPER-2013-064, LHCb-PAPER-2025-042}. Similar studies were conducted with charm mesons by the BABAR experiment~\cite{BaBar:2011ouc}, with strange mesons by the NA62~\cite{NA62:2019eax} and E865~\cite{Appel:2000tc} experiments and with light mesons by the BESIII experiment~\cite{Ablikim_2025}. 

This work investigates the neutrinoless decays \decay{\Bm}{\dordstmumu}, reconstructed through the charm-hadron decays \decay{\Dp}{\Km\pip\pip} and \decay{\Dstarp}{\Dz(\to\Km\pip)\pip}.\footnote{The inclusion of charge-conjugate processes is implied throughout.} A possible Feynman diagram of the processes is shown in Fig.~\ref{fig:0vbb}. Such decays violate lepton number conservation and are forbidden within the SM. 

For context, theoretical predictions for branching fractions of processes of the form \decay{M}{M^\prime\ell\ell}, where $M$ and $M^\prime$ denote initial and final mesons and $\ell\ell$ represent two like-sign leptons of the same generation, have been discussed in Refs.~\cite{Cvetic:2010rw,Abada:2017jjx}, under different assumptions about the Majorana neutrino mass. While the previous study~\cite{LHCb-PAPER-2011-038} considered two possible processes of the decay \decay{\Bm}{\dordstmumu}, this study focuses on a single topology, illustrated in Fig.~\ref{fig:0vbb}, in which the charm meson and the two muons are required to originate from a common vertex. This configuration corresponds to the cases of light ($m_N < m_\pi$) and heavy ($m_N>m_B$) Majorana neutrino discussed in Ref.~\cite{Cvetic:2010rw}. The predicted branching fractions are at the level of $\order (10^{-23})$ and $\order (10^{-22})$, respectively. In contrast, in the intermediate-mass regime ($m_\pi < m_N < m_B$), an on-shell Majorana neutrino would produce a distinct experimental signature involving a displaced vertex, which is not considered in this analysis.

Although these predictions are beyond the reach of present experiments, searches in the \Bm system provide complementary constraints on lepton-number-violating processes and help establish an experimental strategy for future, more sensitive measurements. This analysis performs a model-independent and experimentally clean search for these forbidden processes, focusing on identifying any experimental signature consistent with a neutrinoless \Bm decay.

This study utilizes proton-proton (\proton\proton) collision data, collected with the LHCb detector from 2016 to 2018 at a center-of-mass energy of 13\tev, corresponding to an integrated luminosity of 5.4\invfb. To avoid bias by the experimenter, the results of the analysis were not examined until the full procedure had been finalized. 
This work improves upon a previous study \cite{LHCb-PAPER-2011-038} not only by utilizing a larger dataset but also by carefully estimating backgrounds from $b$-hadron decays in which one or two pions are misidentified as muons. These backgrounds are incorporated in the fit model used to set upper limits on the signal branching fractions. An additional improvement is the use of a boosted decision tree (BDT) classifier~\cite{Breiman} to suppress combinatorial background. For normalization, the decay \decay{\Bm}{\psitwos(\to\jpsi(\to\mup\mun)\pip\pim)\Km} is used, which has the same final-state particles as the signal decays and a high branching fraction. 

The \decay{\Bm}{\dordstarp\mun\mun} branching fractions are determined from the measured yields of signal and normalization decays according to
\begin{equation}\label{eq:signal_BF}
    \BF\left(\decay{\Bm}{\dordstarp\mun\mun}\right) = \dfrac{N_\mathrm{sig}}{N_\mathrm{norm}}\cdot \dfrac{\eps_\mathrm{norm}}{\eps_\mathrm{sig}}\cdot \dfrac{\BF_\mathrm{norm}}{\BF_\dordstarp},
\end{equation}
where $N_\mathrm{sig}$ ($N_\mathrm{norm}$) is the yield of the signal (normalization) decays obtained from an extended unbinned maximum-likelihood fit to the invariant-mass distributions of \Bm candidates, $\eps_\mathrm{sig}$ ($\eps_\mathrm{norm}$) corresponds to the signal (normalization) efficiency estimated with simulation and calibration data, $\BF_\mathrm{norm}$ denotes the known branching fraction of the full decay chain of the normalization channel, and $\BF_\dordstarp$ is the known branching fraction of the subsequent decays of the \dordstarp mesons~\cite{PDG2024}. 

\section{Detector and simulation}

The \lhcb detector~\cite{LHCb-DP-2008-001,LHCb-DP-2014-002} is a single-arm forward
spectrometer covering the \mbox{pseudorapidity} range $2<\eta <5$,
designed for the study of particles containing \bquark or \cquark
quarks. The detector used to collect the data analyzed in this paper includes a high-precision tracking system
consisting of a silicon-strip vertex detector surrounding the $pp$
interaction region, a large-area silicon-strip detector located
upstream of a dipole magnet with a bending power of about
$4{\mathrm{\,T\,m}}$, and three stations of silicon-strip detectors and straw
drift tubes
placed downstream of the magnet.
The tracking system provides a measurement of the momentum, \ptot, of charged particles with
a relative uncertainty that varies from 0.5\% at low momentum to 1.0\% at 200\gev.\footnote{Natural units with $\hbar=c=1$ are used throughout.}
The minimum distance of a track to a primary $pp$ collision vertex (PV), the impact parameter (IP), 
is measured with a resolution of $(15+29/\pt)\mum$,
where \pt is the component of the momentum transverse to the beam, in\,\gev.
Different types of charged hadrons are distinguished using information
from two ring-imaging Cherenkov detectors. 
Photons, electrons, and hadrons are identified by a calorimeter system consisting of
scintillating-pad and preshower detectors and electromagnetic and hadronic calorimeters. Muons are identified by a
system composed of alternating layers of iron and multiwire
proportional chambers.
The online event selection is performed by a trigger, 
which consists of a hardware stage, based on information from the calorimeter and muon
systems, followed by a software stage, which applies a full event
reconstruction.

Simulation plays a crucial role in two key areas: first, to accurately model the effects of the detector acceptance and the applied selection criteria; and second, to characterize the shape of misidentified backgrounds and to estimate their contributions to the data sample. In the simulation, $pp$ collisions are generated using \pythia~\cite{Sjostrand:2007gs,*Sjostrand:2006za} with a specific \lhcb configuration~\cite{LHCb-PROC-2010-056}. Decays of unstable particles are described by \evtgen~\cite{Lange:2001uf}, in which final-state radiation is generated using \photos~\cite{davidson2015photos}. The interaction of the generated particles with the detector, and
 its response, are implemented using the \geant toolkit~\cite{Allison:2006ve, *Agostinelli:2002hh} as described in Ref.~\cite{LHCb-PROC-2011-006}. The simulated samples are corrected to account for known
data-simulation differences in the \B-meson production
kinematics and detector occupancy, as well as particle-identification (PID) performances. Signal decays are simulated uniformly across the \dordstmumu phase space, since no well-established decay model exists. 

\section{Selection}

At the hardware trigger stage, events are required to have at least one muon with high transverse momentum. The software trigger requires a two-, three-, or four-track secondary vertex with a significant displacement from any primary $pp$ interaction vertex. A multivariate algorithm is used for the identification of secondary vertices consistent with the decay of a $b$ hadron. Triggered data further undergo a centralized, offline processing step~\cite{Stripping}. In events selected by the trigger, the \Bm invariant mass is determined using a kinematic fit \cite{Hulsbergen:2005pu}. The fit constrains the invariant mass of the \Km\pip\pip system to the known \Dp (\Dstarp) mass~\cite{PDG2024}. In the \Dstarp channel, the invariant mass of the \Km\pip system is additionally constrained to the known \Dz mass~\cite{PDG2024}. No explicit kaon PID requirement is applied in the signal channels. To suppress combinatorial background, the reconstructed invariant mass of the \Km\pip\pip system is required to lie within $\pm 2.5\sigma$ of the known \Dp (\Dstarp) mass~\cite{PDG2024} mass, where $\sigma$ is the mass resolution determined in simulation. In the \Dstarp channel, the reconstructed \Km\pip invariant mass is similarly required to be within $\pm 2.5\sigma$ of the known \Dz mass.

The \Bm candidates are reconstructed by combining \dordstarp candidates with two same-sign muons. The charm meson and the two muons are required to originate from a common vertex, thereby suppressing contributions from long-lived particles. The invariant mass of the \Bm candidates is required to lie between 5050 and 5500\mev. The mass resolution after the kinematic fit is approximately 38\mev, as determined from simulated signal samples. The chosen window therefore contains the full signal region with sufficient sidebands for background modeling. The tracks of all final-state particles must be inconsistent with originating from any PV, as quantified by the impact-parameter \chisq (\chisqip), which measures the compatibility of a track with a given vertex. Kaon and pion candidates in the signal channels are required to satisfy $\chisqip > 4$, while muon candidates must satisfy $\chisqip > 9$. Compared to the previous LHCb study~\cite{LHCb-PAPER-2011-038}, tighter muon PID requirements are applied to both muon candidates to suppress backgrounds from pions misidentified as muons, which were not explicitly considered in that analysis. The selection retains approximately 83\% of signal, while reducing the pion misidentification probability to below 1\% per track. Backgrounds from \decay{\jpsi}{\mup\mun} and \decay{\psitwos}{\mup\mun} decays, where one of the muons is misidentified and used in the reconstruction of the \dordstarp meson, are suppressed using invariant-mass vetoes. For each muon-pion combination in the event, the invariant mass is computed after assigning the muon mass hypothesis to the pion candidate. Candidates are rejected if any such combination lies within narrow mass windows around the known \jpsi and \psitwos masses\cite{PDG2024}. In the case of the \psitwos veto, this requirement is applied with the pion candidate satisfying the muon PID requirements. The mass-window sizes are determined from studies of the corresponding invariant-mass distributions in data and simulation and are chosen to suppress residual charmonium contributions while  maintaining high signal efficiency.

In the normalization channel, \decay{\Bm}{\psitwos(\to\jpsi(\to\mup\mun)\pip\pim)\Km}, the \Bm candidates are reconstructed using the same selection criteria for the two muons as in the signal channels, except for their charge requirements. The PID requirements, based on differential log-likelihood variables provided by the LHCb PID system~\cite{LHCb-DP-2014-002}, are applied to the kaon and pion candidates, unlike the signal channels. In addition, kaon and pion candidates are required to satisfy a stricter requirement of $\chisqip > 9$ to further suppress combinatorial background. The difference in selection efficiency between the normalization and signal channels is determined using simulation. Since the measurement is performed as a ratio of branching fractions, and the two channels share identical final-state particle content and similar kinematic and topological properties, most reconstruction, trigger, and selection effects cancel in the efficiency ratio. Residual differences are evaluated using simulation and included as systematic uncertainties.

A multivariate selection with a BDT classifier, implemented in the scikit-learn toolkit~\cite{Scikit-learn-paper}, is utilized to separate each signal from the combinatorial background. The BDT classifier is trained and applied independently for the \Dp and \Dstarp decay channels. The signal sample is defined as simulated candidates whose $\dordstarp\mun\mun$ invariant mass lies within $\pm 2.5\sigma$ of the \Bm mass peak. The upper mass sideband, spanning from $2.5\sigma$ above the \Bm mass peak up to 6\gev, is used as the background sample, as it provides a clean sample dominated by random track combinations. The muon PID requirements are not applied in the background sample to ensure a sufficiently large sample and avoid overtraining. 
The variables used in the training of the BDT classifier comprise kinematic and topological quantities, including the transverse momentum of the \Bm, \Km, \pip and muon candidates; the IP and flight distance of the \Bm and \dordstarp particles; and the vertex fit \chisq of the \Bm candidate. The optimal BDT classifier selection is determined using Punzi's figure of merit~\cite{Punzi:2003bu} for a target significance of three standard deviations ($\sigma$).
No dedicated BDT is trained for the selection of the normalization channel. Instead, the BDT classifier developed for the \Dp channel is applied. Although the \Dp and normalization channel do not share identical topologies, the input distributions of the \Dp BDT are more similar to those of the normalization channel than those of the \Dstarp BDT. The BDT requirement in the normalization channel is chosen such that the efficiency, as determined from simulation, is approximately 50\%. This working point provides good agreement between data and simulation while maintaining adequate statistical precision. 

\section{Background sources}
\label{sec:background}

After the BDT selection, which suppresses 99.5\% of the combinatorial background, the dominant remaining background sources are due to misidentified decays such as \decay{\Bm}{\dordstarp\pim\mun\neumb} and \decay{\Bm}{\dordstarp\pim\pim} decays, where one or two pions are reconstructed as muons. The former do not form a peak in the \Bm mass due to the presence of the undetected neutrino in the final state and populate the lower sideband of the selected mass range. The latter appear as a peaking structure near the expected signal peaks. 

Pion misidentification as a muon is often associated with a pion decaying in flight and producing a genuine muon that can satisfy both the hardware trigger and PID requirements. The momentum of such muons is, in general, different from that of the original pion.
To account for these effects, two independent corrections are applied to \decay{\Bm}{\dordstarp\pim\mun\neumb} and \decay{\Bm}{\dordstarp\pim\pim} simulation samples.

The first correction accounts for the pion-to-muon misidentification rate, which is determined using a calibration sample of \decay{\Dstarp}{\Dz(\to\Km\pip)\pip} decays, providing the misidentification probability in bins of the pion momentum, pseudorapidity, and track multiplicity~\cite{LHCb-DP-2018-001}. Simultaneous fits to the \Dz mass ($m_{\Km\pip}$) and the mass difference (\mbox{$\Delta m \equiv m_{\Km\pip\pip} - m_{\Km\pip}$}) are performed to extract the yields of pions that pass or fail the muon-identification requirements. The linear correlation coefficient between the two fit variables is found to be negligible. These yields are used to determine the misidentification rate. 

In the \decay{\Dstarp}{\Dz \pip} candidate sample, the $m_{\Km\pip}$ distribution is modeled by a core component described by two Gaussian probability density functions (PDFs) sharing a common mean but with different widths and a shoulderlike structure modeled by two Gaussian PDFs with distinct means and widths. The shoulder structure primarily accounts for pion-to-muon misidentification of pions decaying in flight to produce genuine muons. The $\Delta m$ distribution is parametrized with the sum of three Gaussian PDFs, two of which share a common mean. 

The fit model contains three main background sources. The first is the misreconstructed \Dstarp background, which arises from a genuine \Dz meson combined with an unrelated pion. Its $m_{\Km\pip}$ distribution is modeled using the same shape as the signal, while its $\Delta m$ distribution is described by a threshold function,  
\begin{equation}
\label{eq:thres_func}
 f(x) = \left(\frac{\Delta m}{a}-1\right)^b e^{-c(\Delta m/a)},   
\end{equation}
where $a$, $b$, and $c$ are floating parameters. The second background source is the misreconstructed \Dz component, dominated by partially reconstructed decays such as \decay{\Dz}{\Km\mup\neum}.  Its $m_{\Km\pip}$ shape is modeled with an exponential PDF, while its $\Delta m$ shape follows that of the signal. Finally, the combinatorial background is described by an exponential function in $m_{\Km\pip}$, and the same threshold function of Eq.~\ref{eq:thres_func} in $\Delta m$.

\begin{figure}[tbp]
    \centering
    \includegraphics[width=\linewidth]{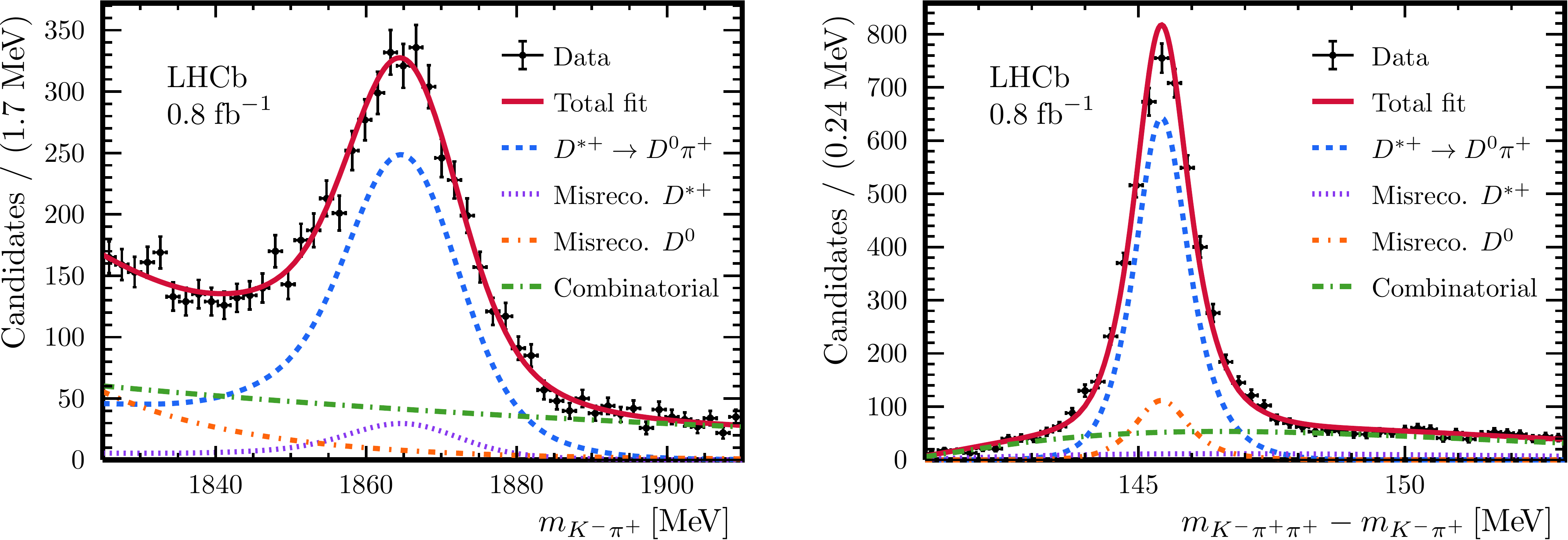}
    \caption{An example of simultaneous fit to (left) the \Dz invariant mass and (right) $\Delta m$ distributions of the \decay{\Dstarp}{\Dz(\to\Km\pip)\pip} candidates that satisfy the muon PID requirement in the bin defined by $p_\pi\in[1.0,\,14.0]\gev$, $\eta\in [1.53,2.83]$, and track multiplicity $\in [0,103]$. This fit is performed separately for different years of data taking and magnetic field polarities, shown here for 2016 data with the magnetic field pointing downwards.}
    \label{fig:custom-pid-2dfit}
\end{figure}

\begin{figure}
    \centering
    \includegraphics[width=0.5\textwidth]{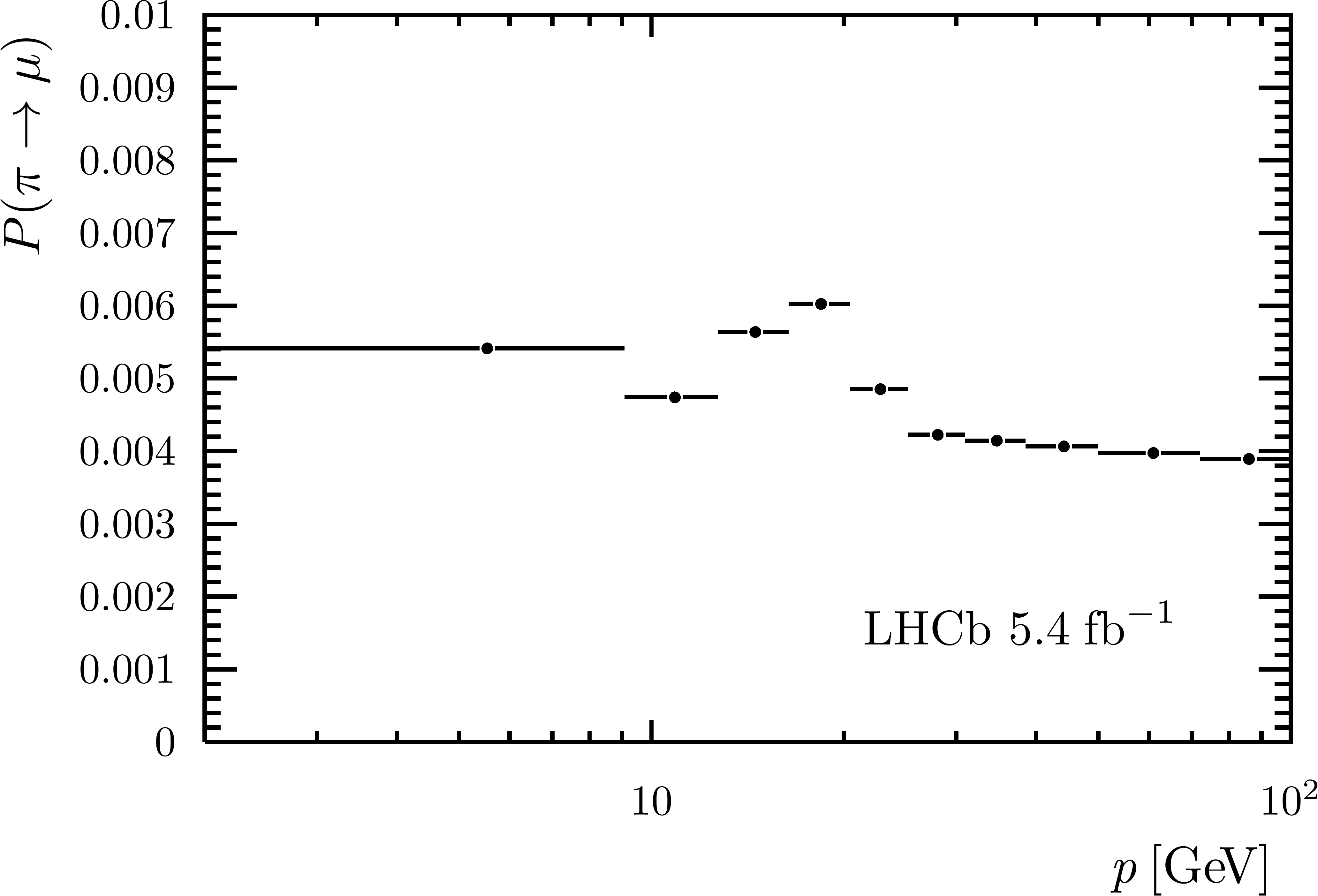}
    \caption{Probability of a pion being misidentified as muon as a function of its momentum.}
    \label{fig:misidrate-p}
\end{figure}

Figure~\ref{fig:custom-pid-2dfit} shows one of the fits to the \decay{\Dstarp}{\Dz(\to\Km\pip)\pip} candidates of the 2016 sample, with pion momentum between 1.0 and 14.0\gev, pion pseudorapidity between 1.53 and 2.83, and track multiplicity of the event smaller than 103. The effect of momentum misreconstruction due to pions decaying in flight is visible in the low-mass tail of the \Dz mass peak. The measured pion-to-muon misidentification rate as a function of the pion momentum  is shown in Fig.~\ref{fig:misidrate-p}.

\begin{figure}[tbp]
    \centering
    \includegraphics[width=0.46\textwidth]{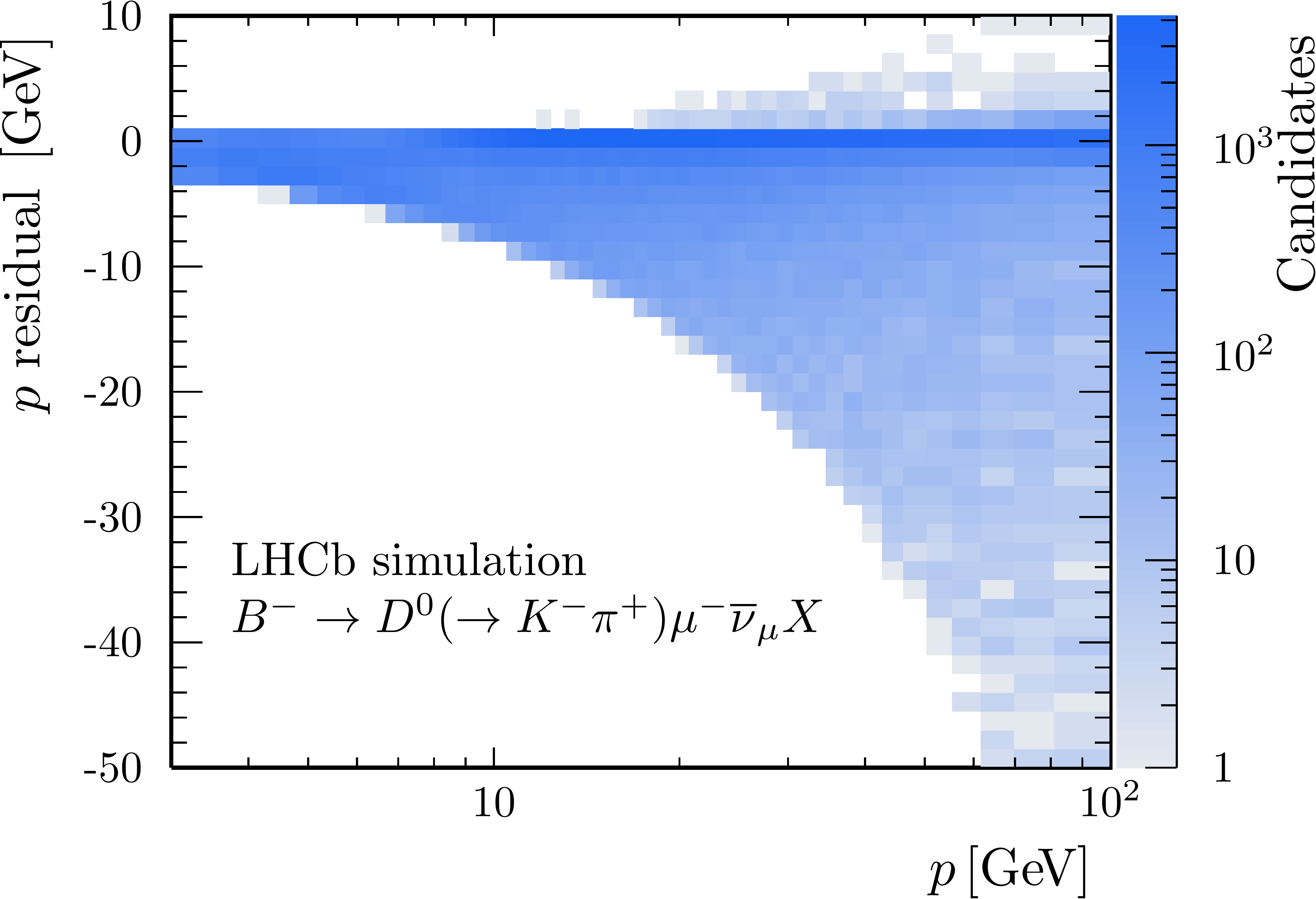}\quad
    \includegraphics[width=0.43\textwidth]{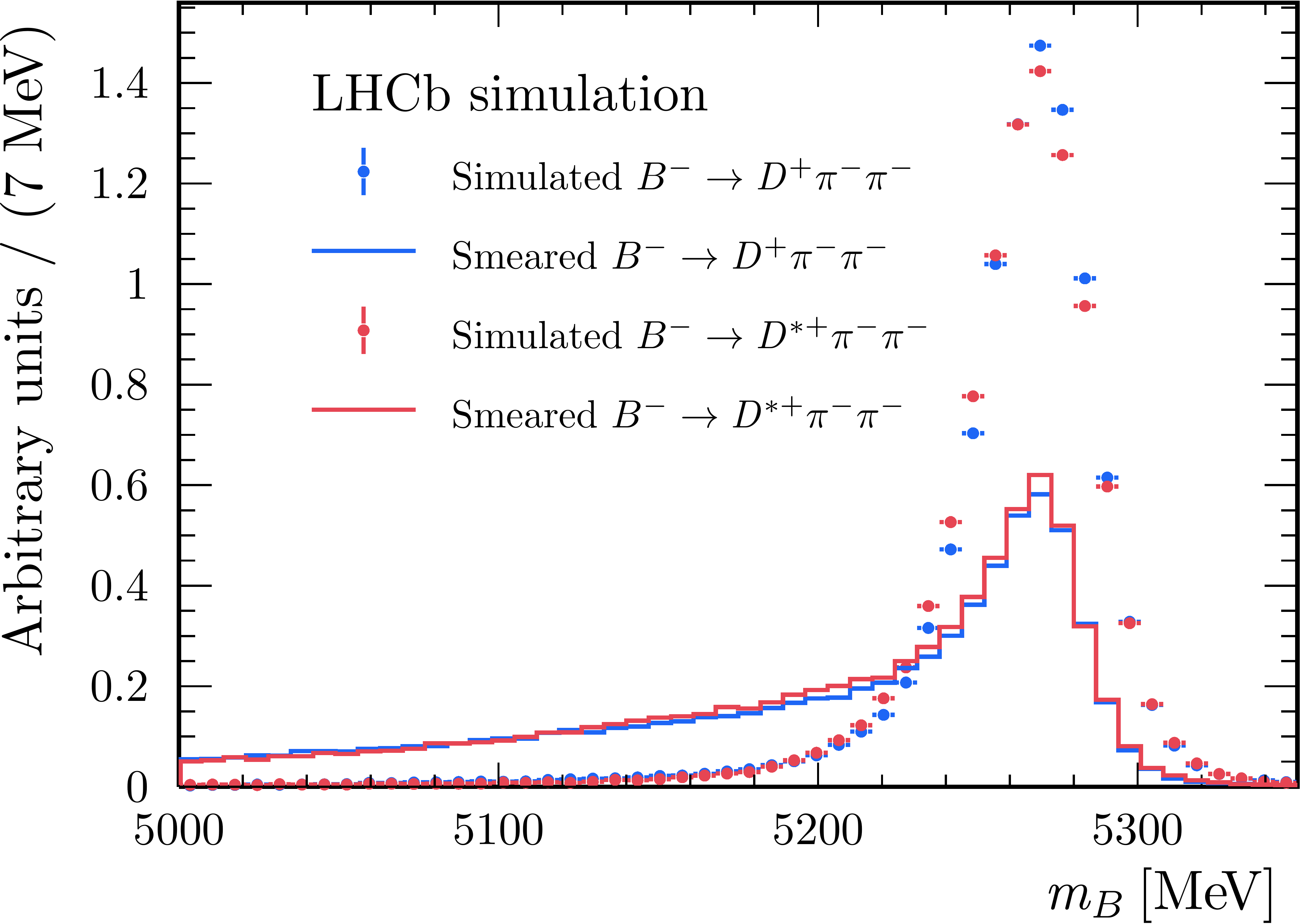}
    \caption{(Left) Pion momentum residual as a function of momentum in the simulated \decay{\Bm}{\Dz\mun\neumb X} sample. (Right) Invariant-mass distribution of the simulated \decay{\Bm}{\dordstarp\pim\pim} candidates misidentified as \decay{\Bm}{\dordstarp\mun\mun}, before and after smearing.}
    \label{fig:pion-decay-in-flight}
\end{figure}

The second correction addresses the effect of momentum degradation arising from pion decays in flight.
Since the momentum of the resulting muon differs from that of the original pion, this effect is studied using simulation.
Available \decay{\Bm}{\dordstarp\pim\mun\neumb} and \decay{\Bm}{\dordstarp\pim\pim} simulation samples do not have sufficient size to show any sensitivity to the aforementioned pion momentum misreconstruction. To address this, a separate, larger simulated sample is used, containing \decay{\Bm}{\Dz\mun\neumb X} decays, where $X$ denotes any additional particles originating from excited charm mesons. Unlike the full simulation, no other particles from the $\proton\proton$ collision are generated in this sample. The pion from \decay{\Dz}{\Km\pip} decays is used to model the momentum misreconstruction. Figure~\ref{fig:pion-decay-in-flight} (left) shows the distributions of the momentum residuals, $p_\mathrm{reco}-p_\mathrm{true}$, of the pion from \decay{\Dz}{\Km\pip} decays in bins of its momentum. The spread toward negative momentum residual reflects the pion decay-in-flight effect, which is more evident at high momenta. The momentum residual distribution is used to smear the reconstructed momentum of pions in the \decay{\Bm}{\dordstarp\pim\mun\neumb} and \decay{\Bm}{\dordstarp\pim\pim} simulation samples, in bins of momentum. The smeared momentum is used to recompute the measured \Bm candidate mass. The comparison of the simulated \decay{\Bm}{\dordstpipi} mass distribution before and after smearing is shown in Fig.~\ref{fig:pion-decay-in-flight} (right). The smeared distribution provides a more accurate description of the shapes of the two background sources in which pions are misidentified as muons due to decay in flight and also reproduces the tail effects observed in the calibration sample shown in Fig.~\ref{fig:custom-pid-2dfit} (left).

Simulated samples of \decay{\Bm}{\dordstarp\pim\mun\neumb} and \decay{\Bm}{\dordstarp\pim\pim} decays are used to estimate, respectively, the yields of singly and doubly misidentified backgrounds in data, where one or two pions are reconstructed as muons. Their selection efficiencies are evaluated by applying the same criteria as for the signal channels, except for the trigger and the PID requirements. The trigger efficiency is taken directly from the signal channels, while the misidentification rate is determined using the data-driven method described above and applied to the simulated sample as an event-by-event weight to candidates reconstructed without muon PID requirements. 

Two sources of systematic uncertainty are considered in the estimation of the yields of misidentified backgrounds. The first originates from the difference between misidentification rates obtained using the simultaneous fit to $m_{\Km\pip}$ and $\Delta m$ and those obtained using the PID calibration procedure described in Ref.~\cite{LHCb-DP-2018-001}. The relative variations are 46\% and 49\% for \decay{\Bm}{\Dp\pim\pim} and \decay{\Bm}{\Dstarp\pim\pim} decays, respectively. The second source of systematic uncertainty is related to the variation of the trigger efficiency across the Dalitz plane, as described in Sec.~\ref{sec:systematics}. Since the trigger efficiencies obtained for the signal decays are also applied to the misidentified backgrounds, relative uncertainties of 6.5\% and 6.9\% are assigned on the misidentified-background yields in the \decay{\Bm}{\Dp\pim\pim} and \decay{\Bm}{\Dstarp\pim\pim} channels, respectively. 
The estimated yields of different misidentified decays are listed in Table~\ref{tab:misid-yields}.

\begin{table}[tbp]
    \centering
    \caption{Estimated yields of backgrounds with one or two pions misidentified as muons. The first uncertainties are due to the limited size of the simulation sample, and the second account for systematic effects.}
    \begin{tabular}{lr@{\:$\pm$\:}l@{\:$\pm$\:}l}
    \toprule
        Decay & \multicolumn{3}{c}{Estimated yield}  \\
        \midrule
        \decay{\Bm}{\Dp\pim\pim} & 7.8 & 0.5& 3.6 \\
\decay{\Bm}{\Dstarp\pim\pim} & 0.8 & 0.2& 0.4 \\
\decay{\Bm}{\Dp\pim\mun\neumb} & 52 & 10& 13 \\
\decay{\Bm}{\Dstarp\pim\mun\neumb} & 12.2 & 4.5& 3.1 \\
        \bottomrule
    \end{tabular}
    \label{tab:misid-yields}
\end{table}

\section{Signal yields}

In the fit to the \Bm invariant-mass distribution, the contributions are modeled with a sum of signal and background components. The signal shapes are modeled by a Hypatia function~\cite{Santos:2013gra}. The misidentified backgrounds are modeled by a double-sided Crystal Ball function~\cite{Skwarnicki:1986xj} for the \decay{\Bm}{\dordstarp\pim\pim} decays, and a kernel density estimator (KDE) function \cite{Cranmer:2000du} for the \decay{\Bm}{\dordstarp\pim\mun\neumb} decays. In the KDE, boundary effects are mitigated by applying symmetric reflection of the data at both edges of the observable range. The smoothing parameter is set to double of the default bandwidth. The combinatorial background is described by an exponential function. 

The PDF parameters of the misidentified background components are determined from simulation, except for their yield, as detailed in Sec.~\ref{sec:background}. The expected yield of each misidentified background is constrained in the fit using a log-normal parametrization, 
 \begin{equation}
    N_\mathrm{misID} = \mu\left(\dfrac{\mu+\sigma}{\mu}\right)^\nu,
\end{equation}
where $\mu$ and $\sigma$ correspond to the estimated yield and its total uncertainty, respectively, with the latter obtained as the quadratic sum of statistical and systematic uncertainties listed in Table~\ref{tab:misid-yields}, and $\nu$ is a nuisance parameter following a standard normal distribution.

\begin{figure}[tbp]
    \centering
    \includegraphics[width=0.49\linewidth]{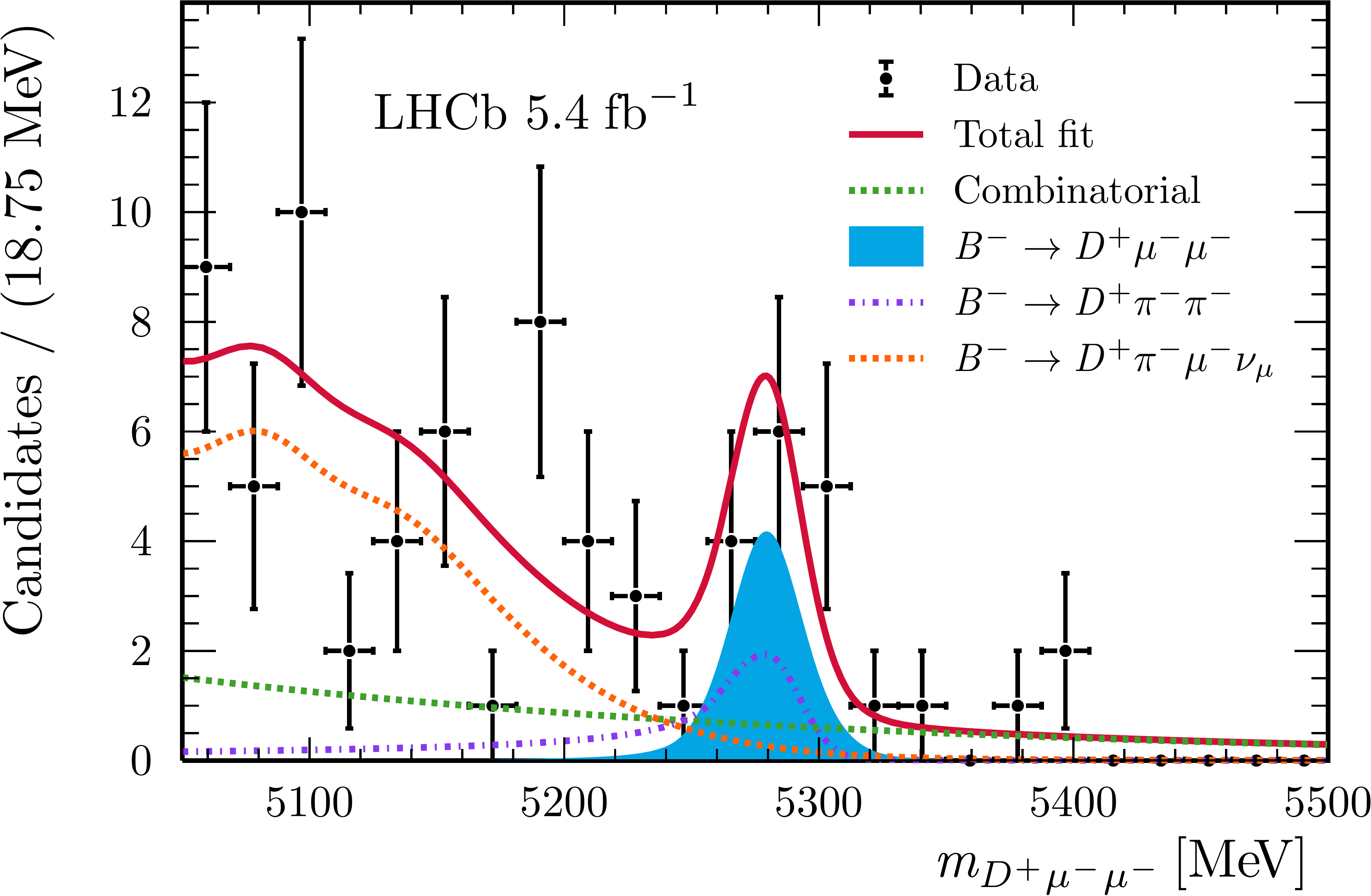}
    \includegraphics[width=0.49\linewidth]{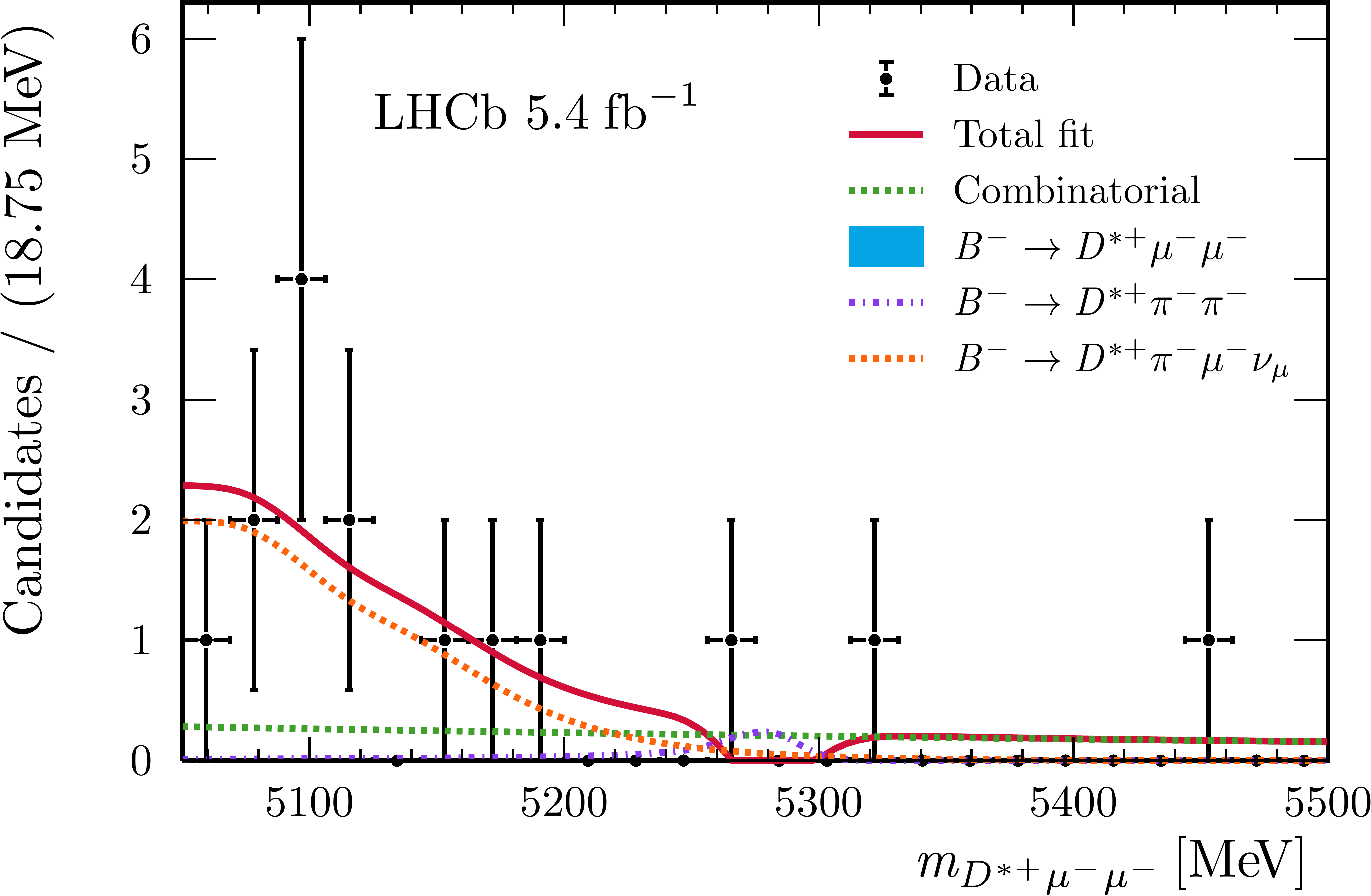}
    \caption{Invariant-mass distribution of (left) \decay{\Bm}{\Dp\mun\mun} and (right) \decay{\Bm}{\Dstarp\mun\mun} candidates with the fit results also shown.}
    \label{fig:fit_dordst_data}
\end{figure}

The fits to the invariant-mass distributions for the \decay{\Bm}{\Dp\mun\mun} and \decay{\Bm}{\Dstarp\mun\mun} decays, as shown in Fig.~\ref{fig:fit_dordst_data}, yield a signal of $9.0 \pm 5.3$ events for the \Dp mode, and $-1.7 \pm 2.1$ events for the \Dstarp mode, respectively.

The normalization yield,  $N_\mathrm{norm}= 24\,460 \pm 160$, is obtained from a fit using a Hypatia function for the normalization channel, and an exponential PDF for the combinatorial background. 

\section{Efficiencies and systematic uncertainties}\label{sec:systematics}

The efficiency terms in Eq.~\ref{eq:signal_BF}, $\eps_\mathrm{sig}$ and $\eps_\mathrm{norm}$, combine the efficiencies of various selections. All efficiencies of signal and normalization channels, except those due to the PID requirements, are determined using simulation. The PID efficiency is evaluated through a data-driven approach \cite{LHCb-DP-2018-001} based on the \sPlot technique \cite{Pivk:2004ty}. The $\eps_\mathrm{norm}/\eps_\mathrm{sig}$ ratio is determined to be $0.36\pm 0.01 \pm 0.06$ for the \Dp channel and $0.69\pm 0.01\pm0.16$ for the \Dstarp channel, where the first uncertainty is statistical and the second is systematic.

The following systematic uncertainties are taken into account when evaluating the efficiency ratio $\eps_\mathrm{norm}/\eps_\mathrm{sig}$, and are summarized in Table~\ref{tab:systematics_eff}. Simulations of signal channels are generated uniformly across phase space. To account for the efficiency differences in a specific new physics model producing the decay, the standard deviation of the efficiency distribution \cite{hepdata.167818} across the Dalitz plane is taken as a systematic uncertainty. This is the dominant systematic uncertainty in the efficiency ratio. 

The uncertainty of the PID efficiency is estimated in two ways: by comparing its baseline value with the simulated PID response, and by changing the binning schemes in the calibration procedure.  The difference in efficiency ratio due to a small variation in the BDT selection criterion is considered as a systematic uncertainty. The large yield of the normalization decay allows the evaluation of the trigger efficiency following  a data-driven approach~\cite{LHCb-PUB-2014-039}, and its difference with the simulation-based efficiency is regarded as a systematic uncertainty. 

The systematic uncertainty due to the limited size of the simulation samples is negligible. The total systematic uncertainty is computed as the quadratic sum of the individual contributions. 

A systematic uncertainty is assigned to the signal yields to account for the choice of PDF used to model the combinatorial background. The baseline model employs an exponential function, while an alternative model uses a second-order Bernstein polynomial. As no clear preference exists between the two parametrizations, an ensemble of background-only pseudoexperiments is generated, and the standard deviation of the resulting differences in fitted signal yields is taken as the absolute systematic uncertainty. The assigned uncertainties are 0.8 and 0.2 candidates for the \Dp and \Dstarp channels, respectively.

\begin{table}[tb]
  \centering
  \caption{Relative systematic uncertainties on the efficiency ratio between the signal and normalization channels. }
\begin{tabular}{l c c}
\toprule
          & \multicolumn{2}{c}{Systematic Uncertainty (\%)}                                         \\
  Source              & $\decay{\Bm}{\Dp\mun\mun}$            & $\decay{\Bm}{\Dstarp\mun\mun}$          \\
\midrule
Decay model     & $16.8$ & $22.9$ \\
PID calibration & $5.6$    & $5.3$    \\
BDT selection         & $3.8$        & $1.3$        \\
Trigger & $0.7$        & $0.7$        \\
\midrule
Total  & $18$ & $24$ \\
\bottomrule
\end{tabular}
\label{tab:systematics_eff}
\end{table}

\section{Results}

\begin{figure}[tbp]
    \centering
    \includegraphics[width=0.49\linewidth]{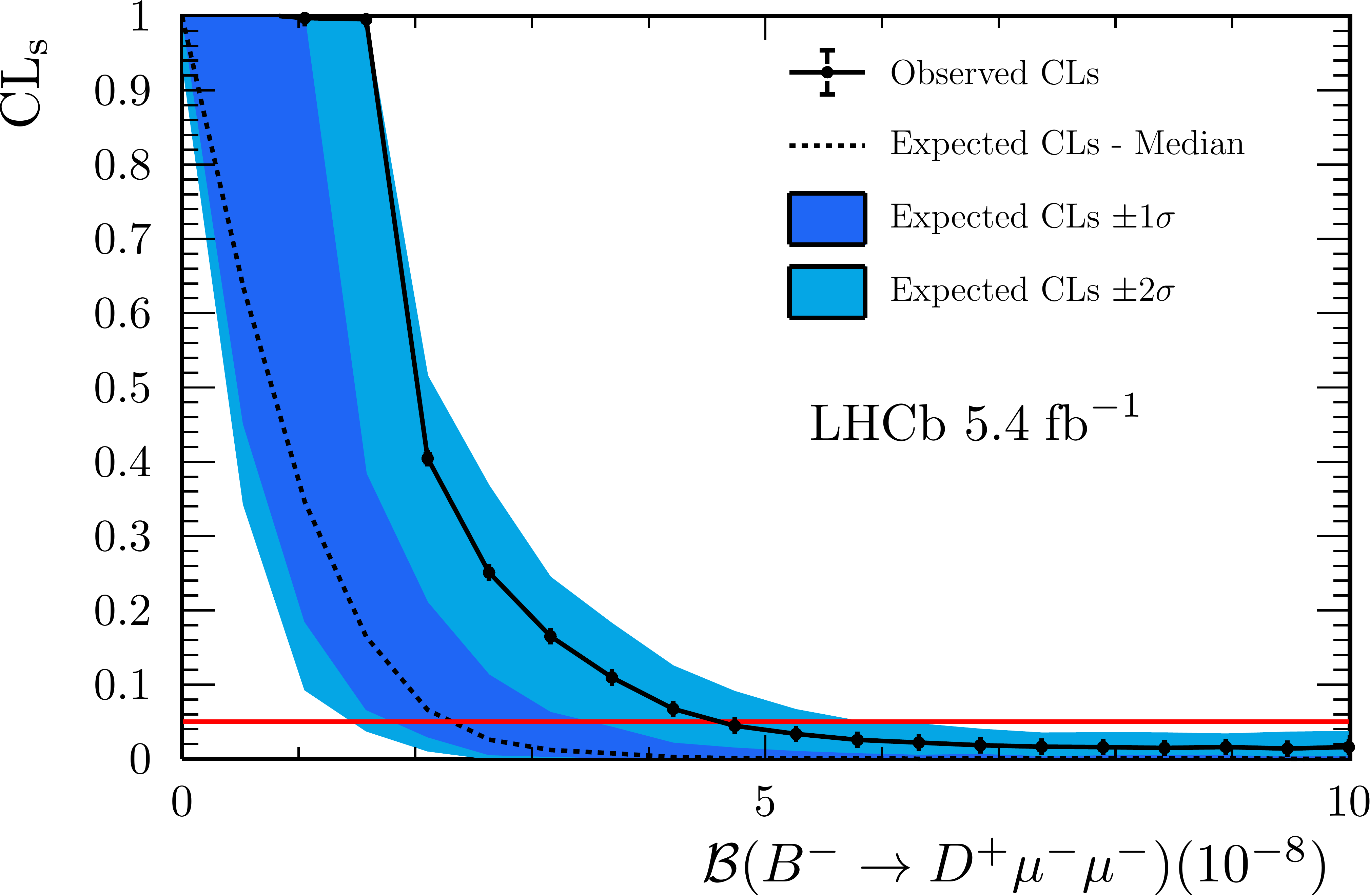}
    \includegraphics[width=0.49\linewidth]{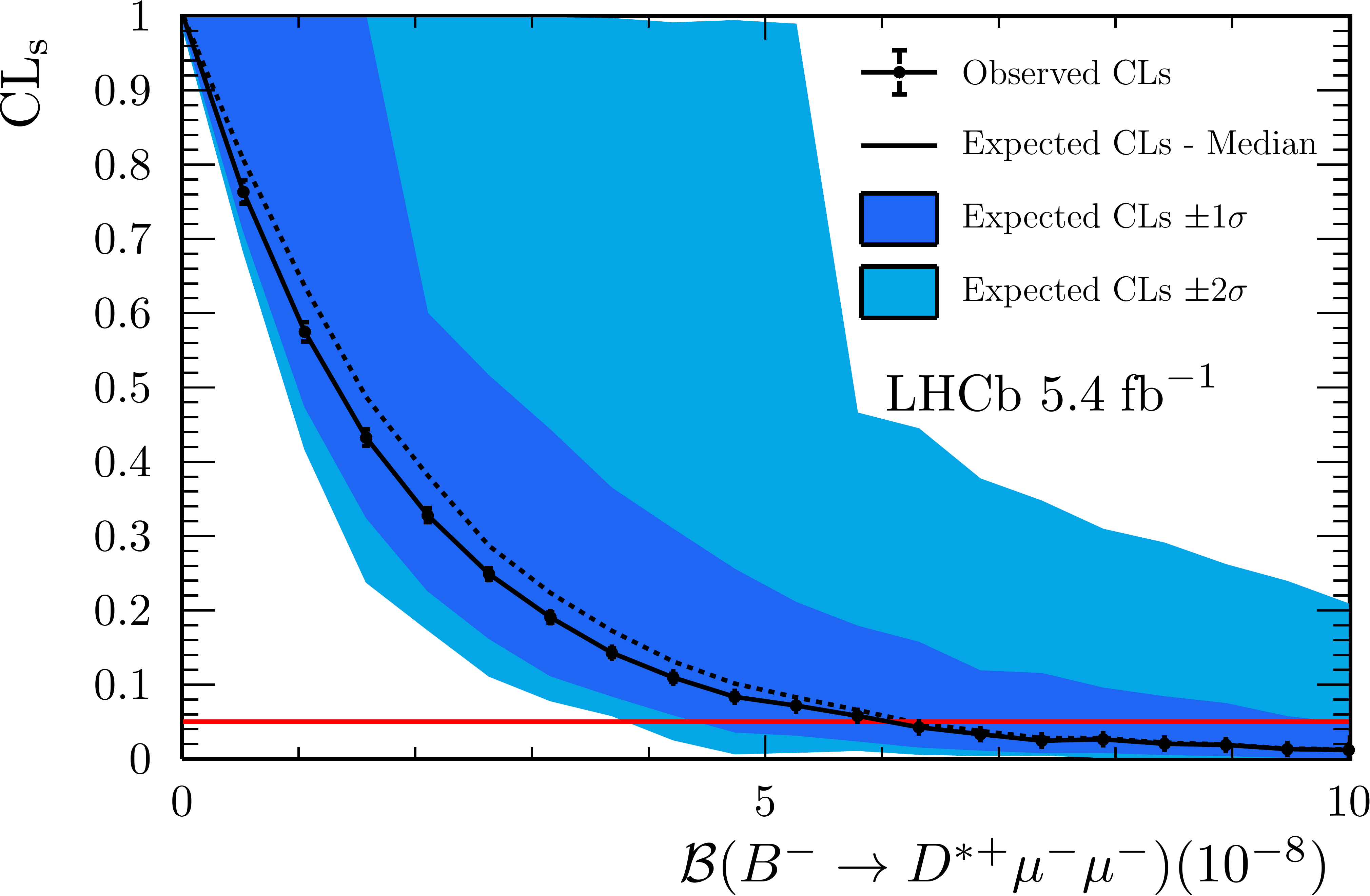}
    \caption{Confidence intervals generated using pseudoexperiments according to the CL$_s$ technique. The horizontal red line indicates $p=0.05$.}
    \label{fig:hti}
\end{figure}

The upper limits of the signal decay branching fractions are computed using the CL$_s$ method \cite{CLs}, implemented in the \roostat project \cite{Moneta:2010pm}. Here CL$_s$ is defined as $(p_{s+b})/(1-p_b)$, where $p_{s+b}$ denotes the $p$-value of the signal-plus-background hypothesis and $p_b$ that of the background-only hypothesis~\cite{PDG2024}. The branching fractions, given by Eq.~\ref{eq:signal_BF}, are evaluated using the signal and normalization yields obtained from the fit and the efficiency ratio determined from simulation. The CL$_s$ method is based on the full fit model, whose likelihood function incorporates the uncertainties on the fit parameters as Gaussian constraints. The upper limits at 90\% (95\%) confidence level~(CL), taken as the points at which the observed CL$_s$ $p$-value dips below $0.10$ ($0.05$) as shown in Fig.~\ref{fig:hti}, are 
\begin{align*}
    \BF (\decay{\Bm}{\Dp\mun\mun}) & < 3.8~(4.6) \times 10^{-8} \text{ at 90\%\;(95\%) CL}, \\
	\BF (\decay{\Bm}{\Dstarp\mun\mun}) & < 4.5~(5.9) \times 10^{-8} \text{ at 90\%\;(95\%) CL}.
\end{align*}

\section{Conclusions}

This updated analysis makes use of a larger dataset and introduces improvements in the estimation of misidentified backgrounds and the suppression of combinatorial background, with respect to the previous analysis~\cite{LHCb-PAPER-2011-038}. These improvements lead to upper limits on the branching fractions that are more than an order of magnitude better than those of previous results~\cite{LHCb-PAPER-2011-038}, highlighting the increased sensitivity and statistical power of the expanded analysis. Although still far from the sensitivity required to probe the theoretical predictions, these results set the most stringent limits to date on the branching fractions of the \decay{\Bm}{\Dp\mun\mun} and \decay{\Bm}{\Dstarp\mun\mun} decays. 
 
\section*{Acknowledgements}
\noindent We express our gratitude to our colleagues in the CERN
accelerator departments for the excellent performance of the LHC. We
thank the technical and administrative staff at the LHCb
institutes.
We acknowledge support from CERN and from the national agencies:
ARC (Australia);
CAPES, CNPq, FAPERJ, and FINEP (Brazil); 
MOST and NSFC (China); 
CNRS/IN2P3 (France); 
BMFTR, DFG, and MPG (Germany);
INFN (Italy); 
NWO (Netherlands); 
MNiSW and NCN (Poland); 
MCID/IFA (Romania); 
MICIU and AEI (Spain);
SNSF and SER (Switzerland); 
NASU (Ukraine); 
STFC (United Kingdom); 
and DOE NP and NSF (USA).
We acknowledge the computing resources that are provided by ARDC (Australia), 
CBPF (Brazil),
CERN, 
IHEP, and LZU (China),
IN2P3 (France), 
KIT and DESY (Germany), 
INFN (Italy), 
SURF (Netherlands),
Polish WLCG (Poland),
IFIN-HH (Romania), 
PIC (Spain), CSCS (Switzerland), 
and GridPP (United Kingdom).
We are indebted to the communities behind the multiple open-source
software packages on which we depend.
Individual groups or members have received support from
Key Research Program of Frontier Sciences of CAS, CAS PIFI, CAS CCEPP, 
Minciencias (Colombia);
EPLANET, Marie Sk\l{}odowska-Curie Actions, ERC and NextGenerationEU (European Union);
A*MIDEX, ANR, IPhU, and Labex P2IO and R\'{e}gion Auvergne-Rh\^{o}ne-Alpes (France);
Alexander-von-Humboldt Foundation (Germany);
ICSC (Italy); 
Severo Ochoa and Mar\'ia de Maeztu Units of Excellence, GVA, XuntaGal, GENCAT, InTalent-Inditex, and Prog.~Atracci\'on Talento CM (Spain);
SRC (Sweden);
and the Leverhulme Trust, the Royal Society, and UKRI (United Kingdom).
 \section*{Data availability}

The data that support the findings of this article are openly available~\cite{hepdata.167818}; embargo periods may apply.

\newpage
\addcontentsline{toc}{section}{References}
\bibliographystyle{LHCb}

\begin{mcitethebibliography}{10}
\mciteSetBstSublistMode{n}
\mciteSetBstMaxWidthForm{subitem}{\alph{mcitesubitemcount})}
\mciteSetBstSublistLabelBeginEnd{\mcitemaxwidthsubitemform\space}
{\relax}{\relax}

\bibitem{Romo-Luque:2025skd}
LEGEND collaboration, C.~Romo-Luque,
  \ifthenelse{\boolean{articletitles}}{\emph{{Results of the LEGEND-200
  experiment in the search for
  0{\ensuremath{\nu}}{\ensuremath{\beta}}{\ensuremath{\beta}} decay}},
  }{}\href{https://doi.org/10.1051/epjconf/202533801002}{EPJ Web Conf.\
  \textbf{338} (2025) 01002},
  \href{http://arxiv.org/abs/2508.18573}{{\normalfont\ttfamily
  arXiv:2508.18573}}\relax
\mciteBstWouldAddEndPuncttrue
\mciteSetBstMidEndSepPunct{\mcitedefaultmidpunct}
{\mcitedefaultendpunct}{\mcitedefaultseppunct}\relax
\EndOfBibitem
\bibitem{EXO-200:2019rkq}
EXO-200 collaboration, G.~Anton {\em et~al.},
  \ifthenelse{\boolean{articletitles}}{\emph{{Search for neutrinoless
  double-$\beta$ decay with the complete EXO-200 dataset}},
  }{}\href{https://doi.org/10.1103/PhysRevLett.123.161802}{Phys.\ Rev.\ Lett.\
  \textbf{123} (2019) 161802},
  \href{http://arxiv.org/abs/1906.02723}{{\normalfont\ttfamily
  arXiv:1906.02723}}\relax
\mciteBstWouldAddEndPuncttrue
\mciteSetBstMidEndSepPunct{\mcitedefaultmidpunct}
{\mcitedefaultendpunct}{\mcitedefaultseppunct}\relax
\EndOfBibitem
\bibitem{BELLE:2011bej}
\belle collaboration, O.~Seon {\em et~al.},
  \ifthenelse{\boolean{articletitles}}{\emph{{Search for
  lepton-number-violating $B^+ \to D^- \ell^+ \ell^{\prime +}$ decays}},
  }{}\href{https://doi.org/10.1103/PhysRevD.84.071106}{Phys.\ Rev.\
  \textbf{D84} (2011) 071106},
  \href{http://arxiv.org/abs/1107.0642}{{\normalfont\ttfamily
  arXiv:1107.0642}}\relax
\mciteBstWouldAddEndPuncttrue
\mciteSetBstMidEndSepPunct{\mcitedefaultmidpunct}
{\mcitedefaultendpunct}{\mcitedefaultseppunct}\relax
\EndOfBibitem
\bibitem{LHCb-PAPER-2011-038}
LHCb collaboration, R.~Aaij {\em et~al.},
  \ifthenelse{\boolean{articletitles}}{\emph{{Searches for Majorana neutrinos
  in $\Bm$ decays}},
  }{}\href{https://doi.org/10.1103/PhysRevD.85.112004}{Phys.\ Rev.\
  \textbf{D85} (2012) 112004},
  \href{http://arxiv.org/abs/1201.5600}{{\normalfont\ttfamily
  arXiv:1201.5600}}\relax
\mciteBstWouldAddEndPuncttrue
\mciteSetBstMidEndSepPunct{\mcitedefaultmidpunct}
{\mcitedefaultendpunct}{\mcitedefaultseppunct}\relax
\EndOfBibitem
\bibitem{LHCb-PAPER-2013-064}
LHCb collaboration, R.~Aaij {\em et~al.},
  \ifthenelse{\boolean{articletitles}}{\emph{{Search for Majorana neutrinos in
  \mbox{\decay{\Bm}{\pip\mun\mun}} decays}},
  }{}\href{https://doi.org/10.1103/PhysRevLett.112.131802}{Phys.\ Rev.\ Lett.\
  \textbf{112} (2014) 131802},
  \href{http://arxiv.org/abs/1401.5361}{{\normalfont\ttfamily
  arXiv:1401.5361}}\relax
\mciteBstWouldAddEndPuncttrue
\mciteSetBstMidEndSepPunct{\mcitedefaultmidpunct}
{\mcitedefaultendpunct}{\mcitedefaultseppunct}\relax
\EndOfBibitem
\bibitem{LHCb-PAPER-2025-042}
LHCb collaboration, R.~Aaij {\em et~al.},
  \ifthenelse{\boolean{articletitles}}{\emph{{Search for heavy neutral leptons
  in $B$-meson decays}},
  }{}\href{https://doi.org/10.1007/JHEP03(2026)178}{{JHEP} \textbf{03} (2026)
  178}, \href{http://arxiv.org/abs/2512.14551}{{\normalfont\ttfamily
  arXiv:2512.14551}}\relax
\mciteBstWouldAddEndPuncttrue
\mciteSetBstMidEndSepPunct{\mcitedefaultmidpunct}
{\mcitedefaultendpunct}{\mcitedefaultseppunct}\relax
\EndOfBibitem
\bibitem{BaBar:2011ouc}
BaBar collaboration, J.~P. Lees {\em et~al.},
  \ifthenelse{\boolean{articletitles}}{\emph{{Searches for rare or forbidden
  semileptonic charm decays}},
  }{}\href{https://doi.org/10.1103/PhysRevD.84.072006}{Phys.\ Rev.\
  \textbf{D84} (2011) 072006},
  \href{http://arxiv.org/abs/1107.4465}{{\normalfont\ttfamily
  arXiv:1107.4465}}\relax
\mciteBstWouldAddEndPuncttrue
\mciteSetBstMidEndSepPunct{\mcitedefaultmidpunct}
{\mcitedefaultendpunct}{\mcitedefaultseppunct}\relax
\EndOfBibitem
\bibitem{NA62:2019eax}
NA62 collaboration, D.~Bryman {\em et~al.},
  \ifthenelse{\boolean{articletitles}}{\emph{{Searches for lepton number
  violating $K^+$ decays}},
  }{}\href{https://doi.org/10.1016/j.physletb.2019.07.041}{Phys.\ Lett.\
  \textbf{B797} (2019) 134794},
  \href{http://arxiv.org/abs/1905.07770}{{\normalfont\ttfamily
  arXiv:1905.07770}}\relax
\mciteBstWouldAddEndPuncttrue
\mciteSetBstMidEndSepPunct{\mcitedefaultmidpunct}
{\mcitedefaultendpunct}{\mcitedefaultseppunct}\relax
\EndOfBibitem
\bibitem{Appel:2000tc}
E865 collaboration, R.~Appel {\em et~al.},
  \ifthenelse{\boolean{articletitles}}{\emph{{Search for lepton flavor
  violation in $K^+$ decays into a charged pion and two leptons}},
  }{}\href{https://doi.org/10.1103/PhysRevLett.85.2877}{Phys.\ Rev.\ Lett.\
  \textbf{85} (2000) 2877},
  \href{http://arxiv.org/abs/hep-ex/0006003}{{\normalfont\ttfamily
  arXiv:hep-ex/0006003}}\relax
\mciteBstWouldAddEndPuncttrue
\mciteSetBstMidEndSepPunct{\mcitedefaultmidpunct}
{\mcitedefaultendpunct}{\mcitedefaultseppunct}\relax
\EndOfBibitem
\bibitem{Ablikim_2025}
M.~Ablikim {\em et~al.}, \ifthenelse{\boolean{articletitles}}{\emph{Search for
  the lepton number violation decay $\omega\to\pi^{+}\pi^{+}e^{-}e^{-} +
  c.c.$}, }{}\href{https://doi.org/10.1088/1674-1137/ade955}{Chin.\ Phys.\
  \textbf{C49} (2025) 103002},
  \href{http://arxiv.org/abs/2504.21539}{{\normalfont\ttfamily
  arXiv:2504.21539}}\relax
\mciteBstWouldAddEndPuncttrue
\mciteSetBstMidEndSepPunct{\mcitedefaultmidpunct}
{\mcitedefaultendpunct}{\mcitedefaultseppunct}\relax
\EndOfBibitem
\bibitem{Cvetic:2010rw}
G.~Cvetic, C.~Dib, S.~K. Kang, and C.~S. Kim,
  \ifthenelse{\boolean{articletitles}}{\emph{{Probing Majorana neutrinos in
  rare $K$ and $D$, $D_s$, $B$, $B_c$ meson decays}},
  }{}\href{https://doi.org/10.1103/PhysRevD.82.053010}{Phys.\ Rev.\
  \textbf{D82} (2010) 053010},
  \href{http://arxiv.org/abs/1005.4282}{{\normalfont\ttfamily
  arXiv:1005.4282}}\relax
\mciteBstWouldAddEndPuncttrue
\mciteSetBstMidEndSepPunct{\mcitedefaultmidpunct}
{\mcitedefaultendpunct}{\mcitedefaultseppunct}\relax
\EndOfBibitem
\bibitem{Abada:2017jjx}
A.~Abada {\em et~al.}, \ifthenelse{\boolean{articletitles}}{\emph{{Effective
  Majorana mass matrix from tau and pseudoscalar meson lepton number violating
  decays}}, }{}\href{https://doi.org/10.1007/JHEP02(2018)169}{JHEP \textbf{02}
  (2018) 169}, \href{http://arxiv.org/abs/1712.03984}{{\normalfont\ttfamily
  arXiv:1712.03984}}\relax
\mciteBstWouldAddEndPuncttrue
\mciteSetBstMidEndSepPunct{\mcitedefaultmidpunct}
{\mcitedefaultendpunct}{\mcitedefaultseppunct}\relax
\EndOfBibitem
\bibitem{Breiman}
L.~Breiman, J.~H. Friedman, R.~A. Olshen, and C.~J. Stone, {\em Classification
  and regression trees}, Wadsworth international group, Belmont, California,
  USA, 1984\relax
\mciteBstWouldAddEndPuncttrue
\mciteSetBstMidEndSepPunct{\mcitedefaultmidpunct}
{\mcitedefaultendpunct}{\mcitedefaultseppunct}\relax
\EndOfBibitem
\bibitem{PDG2024}
Particle Data Group, S.~Navas {\em et~al.},
  \ifthenelse{\boolean{articletitles}}{\emph{{\href{http://pdg.lbl.gov/}{Review
  of particle physics}}},
  }{}\href{https://doi.org/10.1103/PhysRevD.110.030001}{Phys.\ Rev.\
  \textbf{D110} (2024) 030001}\relax
\mciteBstWouldAddEndPuncttrue
\mciteSetBstMidEndSepPunct{\mcitedefaultmidpunct}
{\mcitedefaultendpunct}{\mcitedefaultseppunct}\relax
\EndOfBibitem
\bibitem{LHCb-DP-2008-001}
LHCb collaboration, A.~A. Alves~Jr.\ {\em et~al.},
  \ifthenelse{\boolean{articletitles}}{\emph{{The \lhcb detector at the LHC}},
  }{}\href{https://doi.org/10.1088/1748-0221/3/08/S08005}{JINST \textbf{3}
  (2008) S08005}\relax
\mciteBstWouldAddEndPuncttrue
\mciteSetBstMidEndSepPunct{\mcitedefaultmidpunct}
{\mcitedefaultendpunct}{\mcitedefaultseppunct}\relax
\EndOfBibitem
\bibitem{LHCb-DP-2014-002}
LHCb collaboration, R.~Aaij {\em et~al.},
  \ifthenelse{\boolean{articletitles}}{\emph{{LHCb detector performance}},
  }{}\href{https://doi.org/10.1142/S0217751X15300227}{Int.\ J.\ Mod.\ Phys.\
  \textbf{A30} (2015) 1530022},
  \href{http://arxiv.org/abs/1412.6352}{{\normalfont\ttfamily
  arXiv:1412.6352}}\relax
\mciteBstWouldAddEndPuncttrue
\mciteSetBstMidEndSepPunct{\mcitedefaultmidpunct}
{\mcitedefaultendpunct}{\mcitedefaultseppunct}\relax
\EndOfBibitem
\bibitem{Sjostrand:2007gs}
T.~Sj\"{o}strand, S.~Mrenna, and P.~Skands,
  \ifthenelse{\boolean{articletitles}}{\emph{{A brief introduction to PYTHIA
  8.1}}, }{}\href{https://doi.org/10.1016/j.cpc.2008.01.036}{Comput.\ Phys.\
  Commun.\  \textbf{178} (2008) 852},
  \href{http://arxiv.org/abs/0710.3820}{{\normalfont\ttfamily
  arXiv:0710.3820}}\relax
\mciteBstWouldAddEndPuncttrue
\mciteSetBstMidEndSepPunct{\mcitedefaultmidpunct}
{\mcitedefaultendpunct}{\mcitedefaultseppunct}\relax
\EndOfBibitem
\bibitem{Sjostrand:2006za}
T.~Sj\"{o}strand, S.~Mrenna, and P.~Skands,
  \ifthenelse{\boolean{articletitles}}{\emph{{PYTHIA 6.4 physics and manual}},
  }{}\href{https://doi.org/10.1088/1126-6708/2006/05/026}{JHEP \textbf{05}
  (2006) 026}, \href{http://arxiv.org/abs/hep-ph/0603175}{{\normalfont\ttfamily
  arXiv:hep-ph/0603175}}\relax
\mciteBstWouldAddEndPuncttrue
\mciteSetBstMidEndSepPunct{\mcitedefaultmidpunct}
{\mcitedefaultendpunct}{\mcitedefaultseppunct}\relax
\EndOfBibitem
\bibitem{LHCb-PROC-2010-056}
I.~Belyaev {\em et~al.}, \ifthenelse{\boolean{articletitles}}{\emph{{Handling
  of the generation of primary events in Gauss, the LHCb simulation
  framework}}, }{}\href{https://doi.org/10.1088/1742-6596/331/3/032047}{J.\
  Phys.\ Conf.\ Ser.\  \textbf{331} (2011) 032047}\relax
\mciteBstWouldAddEndPuncttrue
\mciteSetBstMidEndSepPunct{\mcitedefaultmidpunct}
{\mcitedefaultendpunct}{\mcitedefaultseppunct}\relax
\EndOfBibitem
\bibitem{Lange:2001uf}
D.~J. Lange, \ifthenelse{\boolean{articletitles}}{\emph{{The EvtGen particle
  decay simulation package}},
  }{}\href{https://doi.org/10.1016/S0168-9002(01)00089-4}{Nucl.\ Instrum.\
  Meth.\  \textbf{A462} (2001) 152}\relax
\mciteBstWouldAddEndPuncttrue
\mciteSetBstMidEndSepPunct{\mcitedefaultmidpunct}
{\mcitedefaultendpunct}{\mcitedefaultseppunct}\relax
\EndOfBibitem
\bibitem{davidson2015photos}
N.~Davidson, T.~Przedzinski, and Z.~Was,
  \ifthenelse{\boolean{articletitles}}{\emph{{PHOTOS interface in C++:
  Technical and physics documentation}},
  }{}\href{https://doi.org/https://doi.org/10.1016/j.cpc.2015.09.013}{Comput.\
  Phys.\ Commun.\  \textbf{199} (2016) 86},
  \href{http://arxiv.org/abs/1011.0937}{{\normalfont\ttfamily
  arXiv:1011.0937}}\relax
\mciteBstWouldAddEndPuncttrue
\mciteSetBstMidEndSepPunct{\mcitedefaultmidpunct}
{\mcitedefaultendpunct}{\mcitedefaultseppunct}\relax
\EndOfBibitem
\bibitem{Allison:2006ve}
Geant4 collaboration, J.~Allison {\em et~al.},
  \ifthenelse{\boolean{articletitles}}{\emph{{Geant4 developments and
  applications}}, }{}\href{https://doi.org/10.1109/TNS.2006.869826}{IEEE
  Trans.\ Nucl.\ Sci.\  \textbf{53} (2006) 270}\relax
\mciteBstWouldAddEndPuncttrue
\mciteSetBstMidEndSepPunct{\mcitedefaultmidpunct}
{\mcitedefaultendpunct}{\mcitedefaultseppunct}\relax
\EndOfBibitem
\bibitem{Agostinelli:2002hh}
Geant4 collaboration, S.~Agostinelli {\em et~al.},
  \ifthenelse{\boolean{articletitles}}{\emph{{Geant4: A simulation toolkit}},
  }{}\href{https://doi.org/10.1016/S0168-9002(03)01368-8}{Nucl.\ Instrum.\
  Meth.\  \textbf{A506} (2003) 250}\relax
\mciteBstWouldAddEndPuncttrue
\mciteSetBstMidEndSepPunct{\mcitedefaultmidpunct}
{\mcitedefaultendpunct}{\mcitedefaultseppunct}\relax
\EndOfBibitem
\bibitem{LHCb-PROC-2011-006}
M.~Clemencic {\em et~al.}, \ifthenelse{\boolean{articletitles}}{\emph{{The
  \lhcb simulation application, Gauss: Design, evolution and experience}},
  }{}\href{https://doi.org/10.1088/1742-6596/331/3/032023}{J.\ Phys.\ Conf.\
  Ser.\  \textbf{331} (2011) 032023}\relax
\mciteBstWouldAddEndPuncttrue
\mciteSetBstMidEndSepPunct{\mcitedefaultmidpunct}
{\mcitedefaultendpunct}{\mcitedefaultseppunct}\relax
\EndOfBibitem
\bibitem{Stripping}
N.~Grieser {\em et~al.}, \ifthenelse{\boolean{articletitles}}{\emph{{The LHCb
  stripping project: Sustainable legacy data processing for high-energy
  physics}}, }{}\href{https://doi.org/10.1007/s41781-025-00151-6}{Comput.\
  Softw.\ Big.\ Sci.\  \textbf{9} (2025) 21},
  \href{http://arxiv.org/abs/2509.05294}{{\normalfont\ttfamily
  arXiv:2509.05294}}\relax
\mciteBstWouldAddEndPuncttrue
\mciteSetBstMidEndSepPunct{\mcitedefaultmidpunct}
{\mcitedefaultendpunct}{\mcitedefaultseppunct}\relax
\EndOfBibitem
\bibitem{Hulsbergen:2005pu}
W.~D. Hulsbergen, \ifthenelse{\boolean{articletitles}}{\emph{{Decay chain
  fitting with a Kalman filter}},
  }{}\href{https://doi.org/10.1016/j.nima.2005.06.078}{Nucl.\ Instrum.\ Meth.\
  \textbf{A552} (2005) 566},
  \href{http://arxiv.org/abs/physics/0503191}{{\normalfont\ttfamily
  arXiv:physics/0503191}}\relax
\mciteBstWouldAddEndPuncttrue
\mciteSetBstMidEndSepPunct{\mcitedefaultmidpunct}
{\mcitedefaultendpunct}{\mcitedefaultseppunct}\relax
\EndOfBibitem
\bibitem{Scikit-learn-paper}
F.~Pedregosa {\em et~al.},
  \ifthenelse{\boolean{articletitles}}{\emph{{Scikit-learn: Machine learning in
  Python}}, }{}J.\ Mach.\ Learn.\ Res.\  \textbf{12} (2011) 2825,
  \href{http://arxiv.org/abs/1201.0490}{{\normalfont\ttfamily
  arXiv:1201.0490}}, {and online at
  \href{http://scikit-learn.org/stable/}{{\texttt{http://scikit-learn.org/stable/}}}}\relax
\mciteBstWouldAddEndPuncttrue
\mciteSetBstMidEndSepPunct{\mcitedefaultmidpunct}
{\mcitedefaultendpunct}{\mcitedefaultseppunct}\relax
\EndOfBibitem
\bibitem{Punzi:2003bu}
G.~Punzi, \ifthenelse{\boolean{articletitles}}{\emph{{Sensitivity of searches
  for new signals and its optimization}}, }{}eConf \textbf{C030908} (2003)
  MODT002, \href{http://arxiv.org/abs/physics/0308063}{{\normalfont\ttfamily
  arXiv:physics/0308063}}\relax
\mciteBstWouldAddEndPuncttrue
\mciteSetBstMidEndSepPunct{\mcitedefaultmidpunct}
{\mcitedefaultendpunct}{\mcitedefaultseppunct}\relax
\EndOfBibitem
\bibitem{LHCb-DP-2018-001}
R.~Aaij {\em et~al.}, \ifthenelse{\boolean{articletitles}}{\emph{{Selection and
  processing of calibration samples to measure the particle identification
  performance of the LHCb experiment in Run 2}},
  }{}\href{https://doi.org/10.1140/epjti/s40485-019-0050-z}{Eur.\ Phys.\ J.\
  Tech.\ Instr.\  \textbf{6} (2019) 1},
  \href{http://arxiv.org/abs/1803.00824}{{\normalfont\ttfamily
  arXiv:1803.00824}}\relax
\mciteBstWouldAddEndPuncttrue
\mciteSetBstMidEndSepPunct{\mcitedefaultmidpunct}
{\mcitedefaultendpunct}{\mcitedefaultseppunct}\relax
\EndOfBibitem
\bibitem{Santos:2013gra}
D.~Mart{\'\i}nez~Santos and F.~Dupertuis,
  \ifthenelse{\boolean{articletitles}}{\emph{{Mass distributions marginalized
  over per-event errors}},
  }{}\href{https://doi.org/10.1016/j.nima.2014.06.081}{Nucl.\ Instrum.\ Meth.\
  \textbf{A764} (2014) 150},
  \href{http://arxiv.org/abs/1312.5000}{{\normalfont\ttfamily
  arXiv:1312.5000}}\relax
\mciteBstWouldAddEndPuncttrue
\mciteSetBstMidEndSepPunct{\mcitedefaultmidpunct}
{\mcitedefaultendpunct}{\mcitedefaultseppunct}\relax
\EndOfBibitem
\bibitem{Skwarnicki:1986xj}
T.~Skwarnicki, {\em {A study of the radiative cascade transitions between the
  Upsilon-prime and Upsilon resonances}}, PhD thesis, Institute of Nuclear
  Physics, Krakow, 1986,
  {\href{http://inspirehep.net/record/230779/}{DESY-F31-86-02}}\relax
\mciteBstWouldAddEndPuncttrue
\mciteSetBstMidEndSepPunct{\mcitedefaultmidpunct}
{\mcitedefaultendpunct}{\mcitedefaultseppunct}\relax
\EndOfBibitem
\bibitem{Cranmer:2000du}
K.~S. Cranmer, \ifthenelse{\boolean{articletitles}}{\emph{{Kernel estimation in
  high-energy physics}},
  }{}\href{https://doi.org/10.1016/S0010-4655(00)00243-5}{Comput.\ Phys.\
  Commun.\  \textbf{136} (2001) 198},
  \href{http://arxiv.org/abs/hep-ex/0011057}{{\normalfont\ttfamily
  arXiv:hep-ex/0011057}}\relax
\mciteBstWouldAddEndPuncttrue
\mciteSetBstMidEndSepPunct{\mcitedefaultmidpunct}
{\mcitedefaultendpunct}{\mcitedefaultseppunct}\relax
\EndOfBibitem
\bibitem{Pivk:2004ty}
M.~Pivk and F.~R. Le~Diberder,
  \ifthenelse{\boolean{articletitles}}{\emph{{sPlot: A statistical tool to
  unfold data distributions}},
  }{}\href{https://doi.org/10.1016/j.nima.2005.08.106}{Nucl.\ Instrum.\ Meth.\
  \textbf{A555} (2005) 356},
  \href{http://arxiv.org/abs/physics/0402083}{{\normalfont\ttfamily
  arXiv:physics/0402083}}\relax
\mciteBstWouldAddEndPuncttrue
\mciteSetBstMidEndSepPunct{\mcitedefaultmidpunct}
{\mcitedefaultendpunct}{\mcitedefaultseppunct}\relax
\EndOfBibitem
\bibitem{hepdata.167818}
{LHCb Collaboration}, \ifthenelse{\boolean{articletitles}}{\emph{{Search for
  lepton-number-violating $B^-\to D^{(*)+}\mu^-\mu^-$ decays}}, }{} {HEPData
  (collection)}, 2026.
\newblock \url{https://doi.org/10.17182/hepdata.167818}\relax
\mciteBstWouldAddEndPuncttrue
\mciteSetBstMidEndSepPunct{\mcitedefaultmidpunct}
{\mcitedefaultendpunct}{\mcitedefaultseppunct}\relax
\EndOfBibitem
\bibitem{LHCb-PUB-2014-039}
S.~Tolk, J.~Albrecht, F.~Dettori, and A.~Pellegrino,
  \ifthenelse{\boolean{articletitles}}{\emph{{Data driven trigger efficiency
  determination at LHCb}}, }{}
  \href{http://cdsweb.cern.ch/search?p=LHCb-PUB-2014-039&f=reportnumber&action_search=Search&c=LHCb+Notes}
  {LHCb-PUB-2014-039}, 2014\relax
\mciteBstWouldAddEndPuncttrue
\mciteSetBstMidEndSepPunct{\mcitedefaultmidpunct}
{\mcitedefaultendpunct}{\mcitedefaultseppunct}\relax
\EndOfBibitem
\bibitem{CLs}
A.~L. Read, \ifthenelse{\boolean{articletitles}}{\emph{{Presentation of search
  results: The CL$_{\rm s}$ technique}},
  }{}\href{https://doi.org/10.1088/0954-3899/28/10/313}{J.\ Phys.\
  \textbf{G28} (2002) 2693}\relax
\mciteBstWouldAddEndPuncttrue
\mciteSetBstMidEndSepPunct{\mcitedefaultmidpunct}
{\mcitedefaultendpunct}{\mcitedefaultseppunct}\relax
\EndOfBibitem
\bibitem{Moneta:2010pm}
L.~Moneta {\em et~al.}, \ifthenelse{\boolean{articletitles}}{\emph{{The
  RooStats project}}, }{}
  \href{https://doi.org/10.22323/1.093.0057}{\textbf{ACAT2010} 057, 2010},
  \href{http://arxiv.org/abs/1009.1003}{{\normalfont\ttfamily
  arXiv:1009.1003}}\relax
\mciteBstWouldAddEndPuncttrue
\mciteSetBstMidEndSepPunct{\mcitedefaultmidpunct}
{\mcitedefaultendpunct}{\mcitedefaultseppunct}\relax
\EndOfBibitem
\end{mcitethebibliography}
\ifx\mcitethebibliography\mciteundefinedmacro
\PackageError{LHCb.bst}{mciteplus.sty has not been loaded}
{This bibstyle requires the use of the mciteplus package.}\fi
\providecommand{\href}[2]{#2}

\newpage
\centerline
{\large\bf LHCb collaboration}
\begin
{flushleft}
\small
R.~Aaij$^{38}$\lhcborcid{0000-0003-0533-1952},
A.S.W.~Abdelmotteleb$^{57}$\lhcborcid{0000-0001-7905-0542},
C.~Abellan~Beteta$^{51}$\lhcborcid{0009-0009-0869-6798},
F.~Abudin\'en$^{57}$\lhcborcid{0000-0002-6737-3528},
T.~Ackernley$^{61}$\lhcborcid{0000-0002-5951-3498},
A. A. ~Adefisoye$^{69}$\lhcborcid{0000-0003-2448-1550},
B.~Adeva$^{47}$\lhcborcid{0000-0001-9756-3712},
M.~Adinolfi$^{55}$\lhcborcid{0000-0002-1326-1264},
P.~Adlarson$^{85}$\lhcborcid{0000-0001-6280-3851},
C.~Agapopoulou$^{14}$\lhcborcid{0000-0002-2368-0147}
% ...

{\footnotesize \it

% ...
$^{14}$Universit{\'e} Paris-Saclay, CNRS/IN2P3, IJCLab, Orsay, France\\
% ...
$^{38}$Nikhef National Institute for Subatomic Physics, Amsterdam, Netherlands\\
% ...
$^{47}$Instituto Galego de F{\'\i}sica de Altas Enerx{\'\i}as (IGFAE), Universidade de Santiago de Compostela, Santiago de Compostela, Spain\\
% ...
$^{51}$Physik-Institut, Universit{\"a}t Z{\"u}rich, Z{\"u}rich, Switzerland\\
% ...
$^{55}$H.H. Wills Physics Laboratory, University of Bristol, Bristol, United Kingdom\\
% ...
$^{57}$Department of Physics, University of Warwick, Coventry, United Kingdom\\
% ...
$^{60}$School of Physics and Astronomy, University of Glasgow, Glasgow, United Kingdom\\
$^{61}$Oliver Lodge Laboratory, University of Liverpool, Liverpool, United Kingdom\\
% ...
$^{69}$Syracuse University, Syracuse, NY, United States\\
% ...
$^{85}$Department of Physics and Astronomy, Uppsala University, Uppsala, Sweden, associated to $^{60}$\\
% ...
}
\end{flushleft}

\end{document}